\documentclass[12pt]{article}

\usepackage{epsfig,epstopdf,graphicx,latexsym,amsmath,amssymb,cite,url,fullpage,appendix}
\usepackage[us]{datetime}

\usepackage{color}
\definecolor{cardinal}{rgb}{0.6,0,0}
\definecolor{darkgreen}{rgb}{0,0.5,0}
\definecolor{golden}{rgb}{0.92, 0.7, 0}
\definecolor{midnight}{rgb}{0, 0, 0.5}
\definecolor{darkblue}{rgb}{0.2, 0, 0.8}

\usepackage[
	colorlinks=false,
	linkcolor=darkblue,  
	urlcolor=blue,    
	filecolor=blue,     
	citecolor=red,
	linktocpage=true,
	pdfstartview=FitV,
	bookmarksopen=true,
	]{hyperref}

\newcommand{\sect}[1]{\setcounter{equation}{0}\section{#1}}
\renewcommand{\theequation}{\arabic{section}.\arabic{equation}}
\def\be{\begin{equation}}
\def\ee{\end{equation}}
\def\bea{\begin{eqnarray}}
\def\eea{\end{eqnarray}}

\def\CC{\mathbb{C}}

\def\RR{\mathbb{R}}
\def\ZZ{\mathbb{Z}}

\def\cA{{\cal A}}

\def\cC{{\cal C}}

\def\cL{{\cal L}}
\def\cM{{\cal M}}
\def\cN{{\cal N}}
\def\cO{{\cal O}}

\def\cS{{\cal S}}

\def\bfor{{\bf 4}}

\def\a{\alpha}

\def\g{\gamma}

\def\D{\Delta}
\def\e{\varepsilon}

\def\s{\sigma}

\def\r{\rho}

\def\z{\zeta}

\def\ie{\textit{i.e.}}
\def\eg{\textit{e.g.}}
\def\cf{\textit{cf.}}
\def\sir{\textsc{ir}}

\def\sss{\textsc{ss}}
\def\smn{\textsc{mn}}

\def\ker{\textrm{ker}}
\def\dim{\textrm{dim}}
\def\codim{\textrm{codim}}

\def\noi{\noindent}
\def\ds{\displaystyle}

\newcommand{\<}{\langle}
\renewcommand{\>}{\rangle}
\newcommand{\w}{\widetilde}
\newcommand{\del}{\partial}
\newcommand{\dm}{{\partial M}}

\def\nonlin{\varphi}
\def\dnonlin{\varphi_0}
\def\nonlinss{{\varphi_\sss}}
\def\lin{\w{\varphi}}
\def\dlin{\w\varphi_0}
\def\dlinext{\w{\varphi}_0^{\textsc{m}}}
\def\surj{\xi}
\def\surjset{\Xi}
\def\PDO{\Psi}
\def\Bound{\Pi}

\date{\today}

\begin{document}  
\begin{titlepage}

\vbox{}
\vbox{}
\vbox{}
\vbox{}
\begin{center}
{\bf\LARGE{Holographic Uniformization}}\vskip2.25cm{
Michael T. Anderson\footnote{\texttt{anderson@math.sunysb.edu}},
Christopher Beem\footnote{\texttt{cbeem@scgp.stonybrook.edu}},
Nikolay Bobev\footnote{\texttt{nbobev@scgp.stonybrook.edu}},
Leonardo Rastelli\footnote{\texttt{leonardo.rastelli@stonybrook.edu}}
}\vskip1cm 
{\textit{$^{1}$Department of Mathematics, Stony Brook University}\\
{\textit{Stony Brook, NY 11794-3651}}}\\
\vskip.3cm
{\textit{$^{2,3}$Simons Center for Geometry and Physics}\\
{\textit{Stony Brook, NY 11794-3636}}}\\
\vskip.3cm
{\textit{$^{4}$C.~N.~Yang Institute for Theoretical Physics, Stony Brook University}\\
{\textit{Stony Brook, NY 11794-3840}}}
\end{center}
\vskip1.25cm

\begin{abstract}
\vskip.15cm
\noindent
We derive and study supergravity BPS flow equations for M5 or D3 branes wrapping 
a Riemann surface. They take the form of novel
geometric flows intrinsically defined on the surface. Their dual field-theoretic interpretation
suggests the existence of solutions interpolating between an arbitrary metric in the UV and the 
constant-curvature metric in the IR. We confirm this conjecture with a rigorous global existence 
proof.
\end{abstract}
\vfill
\today
\end{titlepage}

\tableofcontents
\setcounter{footnote}{0}

\sect{Introduction}\label{sec:intro}

Geometric flow equations are a central subject in modern differential geometry and topology. They also arise  naturally in quantum field theory as  renormalization group (RG) equations in theories whose coupling space  is parametrized by a  Riemannian manifold. A prototypical example is Ricci flow \cite{RicciFlowHamilton,RicciFlow1}, which independently appeared in quantum field theory (in Friedan's thesis \cite{Friedan}) just before being introduced by Hamilton as a tool to attack the geometrization conjecture for three-manifolds \cite{RicciFlowIntroduced}. Ricci flow describes the one-loop RG evolution for the metric of the target manifold ${\cal M}$  of a two-dimensional sigma-model. Under certain assumptions, and after appropriate rescaling, solutions of  Ricci flow tend to a constant curvature metric on ${\cal M}$. Physically, this canonical metric is interpreted as an infrared (IR) stable fixed point; the metric moduli are irrelevant in the RG sense, and they are washed out by the flow.

Here we introduce and study a new class of geometric flows which arise  as holographic  BPS flows for certain supersymmetric large $N$ field theories. We restrict to flows defined on a closed Riemann surface $\cC$;
the very interesting extension to  three-manifolds will be presented elsewhere \cite{inprogress3}. The dual interpretation of the flows  as field-theory RG flows  suggests that they should uniformize the surface, that is, for fixed complex structure on $\cC$ there should exist a solution interpolating between an arbitrary metric on $\cC$ in the ultraviolet (UV) and the attractor metric of  constant  curvature in the IR.   We  confirm this expectation by rigorous mathematical argument.
 
We emphasize from the outset that our flow equations, while certainly related to the physics of renormalization, have a rather different flavor from flows, such as Ricci flow, that admit a more direct field-theoretic RG interpretation.   Indeed  our flow equations are  second-order (elliptic) in RG time, rather than first-order (parabolic) and we study  them as  a boundary-value problem with prescribed UV and IR behavior. This is a familiar predicament.
Quite generally, if one regards  supergravity flow equations as defining an initial-value problem, one needs to constrain the UV data such that the evolution does not lead to unphysical singularities. This is very difficult, and in practice it is more convenient to study instead a boundary value problem with specified UV and IR data. However, this is not in the spirit of the Wilsonian RG, where for all initial UV data there is a well-defined physical flow.  

Our initial motivation comes from physics. We want to test a crucial assumption of the beautiful recent work on four-dimensional $\cN=2$ supersymmetric quantum field theories ``of class $\cS$'' \cite{Gaiotto:2009we,Gaiotto:2008cd,Gaiotto:2009hg}. These are the theories conjectured to arise by compactification on a Riemann surface $\cC$ of the famous six-dimensional $(2,0)$ superconformal field theory (SCFT). 
The appropriate partial topological twist ensures that  $\cN=2$ supersymmetry is preserved in the four non-compact dimensions for arbitrary metric on $\cC$. Then in the IR the theory must flow to a four-dimensional $\cN=2$ SCFT. The complex structure moduli space of $\cC$ is identified with the space of exactly marginal couplings of the four-dimensional SCFT, but the conformal factor of the metric is believed to be RG-irrelevant and thus forgotten in the IR. This is the assumption that we set out to check. 
 
As a Lagrangian description of the $(2,0)$ theory is presently lacking, we do not know how to approach this question in general. Fortunately,  a simplification occurs for large $N$, where $N$ is the rank of the Lie algebra $A_{N-1}$ that characterizes the $(2,0)$ theory. In this limit we can appeal to the $AdS$/CFT correspondence, which states that the $(2,0)$ $A_{N-1}$ theory is dual to eleven-dimensional supergravity in an $AdS_7 \times S^4$ background. In fact for our purposes, it is sufficient to consider the consistent truncation of eleven-dimensional supergravity to seven-dimensional gauged supergravity.\footnote{This truncation necessitates the restriction that $\cC$ be closed. While there is a rich generalization to punctured surfaces \cite{Gaiotto:2009gz}, it is technically much simpler for us to study to the case with no punctures. We also generally assume that $\cC$ has genus ${\bf g} > 1$. This is a less essential restriction: the equations we derive actually describe the low-genus cases as well, though the corresponding flows are singular.} Then the hypothesis  that we would like to check can be rephrased in the language of the ``holographic RG''. One singles out a radial coordinate to play the role of RG time and writes the supergravity BPS equations as evolution equations with respect to this coordinate. The solutions of interest interpolate between an asymptotically locally $AdS_7$ background in the UV and the background $AdS_5 \times \cC$  (where  $\cC$ has fixed constant negative curvature) in the IR. The expectation is that such a solution exists for {\it arbitrary} choice of UV metric on $\cC$.
 
At first sight, the supergravity BPS equations look like a complicated coupled system, but remarkably they can be reduced to the very elegant equation (\ref{M5N=2Heaven}):

\be
\del_\r^2e^\Phi+(\del_x^2+\del_y^2)\Phi=m^2e^\Phi~.\notag
\ee
This is a single elliptic flow equation for a scalar field $\Phi$ intrinsically defined on the surface! In terms of the original  variables,  $\Phi$  is a linear combination of the conformal factor of the metric on $\cC$  and of one of the scalars fields of the gauged supergravity. We can also think of it more covariantly as an equation for an auxiliary metric on $\cC$, of which $\Phi$ is the conformal factor, see (\ref{ABEqM52}). The equation admits an exact solution, which equates  to the previously known solution where $\cC$ is taken to have constant curvature throughout the flow \cite{Maldacena:2000mw}. Linearizing around this constant-curvature flow, it is easy to demonstrate that for infinitesimal perturbations of the UV metric there is always a solution flowing to the attractive fixed point in the IR. Much less trivially, we are able to give a rigorous global existence proof. The proof is based on degree theoretic techniques used in proving existence results for nonlinear elliptic equations. For a survey of this area of nonlinear functional or global analysis, see \cite{N1}. Such methods can be used, for instance, to give a relatively simple proof of the uniformization theorem for surfaces of higher genus \cite{Kazdan}. The proof here is more difficult, since it involves flows with substantially different behaviors in the UV and IR.

We perform a similar analysis for a few other cases of physical interest. The first variation on our theme is  to consider a different partial topological twist of the $(2,0)$ theory compactified on $\cC$, such that only $\cN=1$ supersymmetry is preserved in four dimensions.  In fact there is a whole family of possible twists that preserve $\cN=1$ supersymmetry, and here we restrict to the simplest case, already discussed in \cite{Maldacena:2000mw}; a more comprehensive discussion will appear elsewhere \cite{inprogresstwists}. Another variation is to take as the starting point $\cN=4$ super Yang-Mills, a four-dimensional SCFT, rather than the six-dimensional $(2,0)$ theory. We consider  compactifications of $\cN=4$ SYM on $\cC$ with partial topological twists that preserve either $(4,4)$ or $ (2,2)$ supersymmetry in the two non-compact dimensions. For all of these cases, the holographic RG equations reduce to a single scalar equation on ${\cal C}$.
 
The example of the  (4,4) twist of $\cN=4$ SYM is somewhat special, since one does not expect the IR theory to have a well-defined vacuum state \cite{Maldacena:2000mw}, and correspondingly one finds no $AdS_3 \times {\cal C}$ solution in the dual supergravity. On the other hand, both the $(2,0)$ theory with $\cN =1$ twist and $\cN = 4$ SYM with the $(2,2)$ twist flow in the IR to SCFTs in four and two dimensions, respectively. As before, the field-theoretic expectation is that memory of the UV metric on ${\cal C}$ is lost in the IR. This is confirmed by the analysis of the corresponding scalar flow equations (\ref{M51/4Heaven}) and (\ref{D3N=22Heaven}) which, despite looking less elegant than \eqref{M5N=2Heaven}, have very similar behavior. 

The organization of the paper is as follows. In Section \ref{sec:background}, we review the construction of the field theories of interest by partial twisting of maximally supersymmetric theories. We then recall the realization of these field theories on the worldvolumes of D3 and M5 branes wrapping supersymmetric cycles in Calabi-Yau manifolds. In Sections \ref{sec:M5s} and \ref{sec:D3s}, we go about finding the gravity duals to the partially twisted $(2,0)$ and $\cN=4$ SYM field theories, respectively, and reduce the problem in each case to a single elliptic geometric flow equation on the Riemann surface. We also perform a linearized analysis of these flow equations, interpret the results using AdS/CFT and show that the constant curvature metric on the Riemann surface is a local IR attractor of the flow equations. In Section \ref{sec:theproof}, we provide a global proof that the geometric flows in question uniformize any metric on the Riemann surface for a correct choice of additional boundary data. We further explore the flow of the area of $\cC$ with respect to the auxiliary metric, and find that it decreases monotonically. Many technical details of the computations are reported in the appendices.

\sect{Field Theory, Branes, Supergravity}\label{sec:background}

We begin by reviewing the field theories of interest, their realization on the worldvolumes of M5 and D3 branes, and our approach to constructing their gravity duals. The bulk of the material in this section has appeared previously, in particular in \cite{Maldacena:2000mw} (see also \cite{Fayyazuddin:2000em,Brinne:2000fh}). However, as the analysis in the present work is somewhat more involved than that of \cite{Maldacena:2000mw},  we place special emphasis on symmetries as the basic guiding principle: the symmetries of the partially-twisted field theory can be used to systematically determine the geometry of the brane construction, which in turn completely fixes the Ansatz for the supergravity analysis.

\subsection{Partially twisted field theories}\label{ssec:twist}

We study the $(2,0)$ theory of $A_{N-1}$ type in $d=6$ dimensions and $\cN=4$ SYM with $SU(N)$ gauge group in $d=4$ dimensions, defined on a spacetime of the form
\be
\RR^{1,d-3}\times\cC~,\label{spacetime}
\ee
with $\cC$ a compact Riemann surface of genus ${\bf g} > 1$. Supersymmetry would normally be broken explicitly and completely by the curved background due to the absence of covariantly constant spinors. This situation can be remedied if the theory is (partially) twisted \cite{Witten:1988ze,Bershadsky:1995qy}. Because we consider geometries with product metrics where only a two-dimensional factor is curved, the structure group of the spacetime manifold naturally reduces according to
\be
SO(1,d-1)\to SO(1,d-3)\times SO(2)_{\cC}~.\label{structuregroup}
\ee
A choice of twist is a choice of  Abelian subgroup $SO(2)_{\cC}^\prime\subset SO(2)_{\cC}\times G_R$, with $G_R$ the R-symmetry group of the $d$-dimensional field theory, such that some of the supercharges are invariant under $SO(2)_{\cC}^\prime$. For the theories at hand, the R-symmetry group $G_R$ is $SO(5)$ or $SO(6)$ and there exist a number of inequivalent ways to choose the group $SO(2)_{\cC}^\prime$ so that some supersymmetry is preserved. We restrict our attention to two twists for each theory. We now review these twists and mention some standard facts about the resulting $(d-2)$-dimensional theories.

\subsubsection{Twists of the $(2,0)$ SCFT in six dimensions}\label{sssec:M5twist}

The Poincar\'e supercharges of the $(2,0)$ superconformal algebra transform in the $\bfor\otimes\bfor$ of the maximal bosonic subgroup $SO(1,5)\times SO(5)_R$ and respect a symplectic-Majorana constraint. Because only an Abelian factor of the structure group is being twisted, it is sufficient to consider the maximal torus of the R-symmetry group, $SO(5)_R$. In particular, if we think of $SO(5)_R$ as rotations of $\RR^5_{x_{1-5}}$, then we define $U(1)_{R,12}\times U(1)_{R,34}\subset SO(5)_R$ as the subgroups which rotate the $(x_1,x_2)$ and $(x_3,x_4)$ planes independently. Under the subgroup $SO(1,3)\times SO(2)_{\cC}\times U(1)_{R,12}\times U(1)_{R,34}\subset SO(1,5)\times SO(5)_R$, the supercharges decompose as
\be
\bfor\otimes\bfor\to\Bigl[{\bf(2,1)}_{\textstyle\tfrac{1}{2}}\oplus{\bf(1,2)}_{-\textstyle\tfrac{1}{2}}\Bigr]\otimes\Bigl[\textstyle(\tfrac{1}{2},\tfrac{1}{2})\oplus(-\tfrac{1}{2},\tfrac{1}{2})\oplus(\tfrac{1}{2},-\tfrac{1}{2})\oplus(-\tfrac{1}{2},-\tfrac{1}{2})\Bigr]~,
\label{M5spinordecomp}
\ee
and satisfy a reality constraint coming from the symplectic-Majorana condition. Thus, under a $U(1)$ subgroup generated by a Lie algebra element $t^\prime=t_{\cC}+a t_{12}+ b t_{34}$ (the $t$'s on the right-hand side being the generators of $SO(2)_{\cC}$, $U(1)_{12}$, and $U(1)_{34}$, respectively), the supercharges transform with charges $\pm\frac{1}{2}\pm\frac{a}{2}\pm\frac{b}{2}$. For any choice of $a$ and $b$ such that $a \pm b= \pm1$ there are at least four real, invariant supercharges, so at low energies the theory enjoys four-dimensional $\cN=1$ supersymmetry. In the special case when either $a$ or $b$ is zero, the supersymmetry is enhanced to $\cN=2$ in four dimensions.\footnote{The discussion of twisting here is purely local. In particular, when the twisted theory is defined on a curved background with non-trivial topology there are global obstructions to the procedure outlined except at discrete values of $a$ and $b$. This becomes manifest in Section \ref{ssec:branegeometry}, where the obstructions are geometrized.}

The first twist studied corresponds to the choice $a=1$ and $b=0$. We refer to this as the ``$1/2$ BPS twist''. It has been argued in \cite{Gaiotto:2009we} that these twisted compactifications of the $(2,0)$ theory flow to four-dimensional SCFTs of class $\cS$ \cite{Gaiotto:2008cd, Gaiotto:2009hg}. One key aspect of any theory of class $\cS$ is that it has a moduli space which is equivalent to the complex structure moduli space of an associated Riemann surface -- the ``UV curve''. In \cite{Gaiotto:2009we}, the UV curve was identified with the Riemann surface $\cC$ on which the $(2,0)$ theory is compactified, and it was conjectured that under the subsequent RG flow to a four-dimensional fixed point, all metric data for $\cC$ except for the complex structure are irrelevant. The arguments for this picture are compelling. For example, the space of marginal deformations in the four-dimensional theory leaves no room for additional geometric degrees of freedom, and BPS quantities in the twisted six-dimensional theory are determined by the complex structure alone. Nevertheless, the hard-boiled skeptic cannot rule out the existence of disconnected components in space of IR fixed points.

The second twist considered corresponds to the choice $a=b=1/2$, which is the ``$1/4$ BPS twist''. These theories have been considered in \cite{Benini:2009mz}, where they were identified as the end point of an RG flow triggered by a mass deformation of the $\cN=2$ theory of class $\cS$ for the same UV curve. It was further argued that the moduli space of these theories is the combined space of complex structures and flat $SU(2)$ bundles on the UV curve. Locally, this moduli space is just the product of the complex structure moduli space with the space of $SU(2)$ Wilson lines for the UV curve.

\subsubsection{Twists of $\cN=4$ SYM in four dimensions}\label{sssec:D3twist}

The Poincar\'e supercharges of $\cN=4$ SYM transform in the $[({\bf 2},{\bf 1})\,\oplus\,({\bf 1},{\bf 2})]\,\otimes\,\bfor$ of $SO(1,3)\times SU(4)_R$ with a Majorana constraint. As in the case of the $(2,0)$ theory, it is sufficient to consider a maximal torus of $SU(4)_R \cong SO(6)_R$, which we regard as independent rotations of three planes in $\RR^6_{x_{1-6}}$. Under the subgroup $SO(1,1)\times SO(2)_{\cC}\times U(1)_{12}\times U(1)_{34}\times U(1)_{56}$, the supercharges decompose as
\be
[({\bf 2},{\bf 1})\,\oplus\,({\bf 1},{\bf 2})]\,\otimes\,\bfor\to\Bigl[\textstyle(\pm\frac{1}{2},\pm\frac{1}{2})\Bigr] \otimes \Bigl[\textstyle(\frac{1}{2},\frac{1}{2},\frac{1}{2})\oplus(-\frac
{1}{2},-\frac{1}{2},\frac{1}{2})\oplus(-\frac{1}{2},\frac{1}{2},-\frac{1}{2})\oplus(\frac{1}{2},-\frac{1}{2},-\frac{1}{2})
\Bigr]~.  \label{D3spinordecomp}
\ee
If we consider the $U(1)$ subgroup generated by a Lie algebra element $t^\prime=t_{\cC}+a\,t_{12}+b\,t_{34}+c\,t_{56}$, it is straightforward to check that at least two real supercharges are invariant for $a\pm b\pm c=\pm1$. This is enhanced to four invariant supercharges if $a$, $b$, or $c$ vanish, and eight invariant supercharges if only one of $a$, $b$, and $c$ is non-zero. These classes of twists give rise to theories which flow to two-dimensional theories preserving $\cN=(1,1)$, $\cN=(2,2)$, and $\cN=(4,4)$ supersymmetry, respectively. We focus on the two latter cases as the additional supersymmetry leads to nice simplifications.

First we consider the  ``1/2 BPS twist'' with $(a,b,c)=(0,0,1)$. Since $\cN=4$ SYM has a Lagrangian description, the resulting twisted field theory can be studied quite explicitly, and in \cite{Bershadsky:1995vm} it was argued that the IR fixed point is a sigma model with target space the hyper-K\"ahler moduli space $\cM^H(\cC)$ of solutions to the Hitchin equations. This sigma model explicitly depends only on the complex structure on $\cC$, and so is insensitive to the conformal factor of the metric. Then we study the  ``$1/4$ BPS twist'' with $(a,b,c)=(\frac{1}{2},\frac{1}{2},0)$. This is related to the Donaldson-Witten twist of $\cN=2$ theories in four dimensions where $\cN=4$ SYM is treated as an $\cN=2$ theory with an adjoint hypermultiplet.

\subsection{Brane realization}\label{ssec:branegeometry}

The maximally supersymmetric theories of interest -- \ie, the $A_{N-1}$ $(2,0)$ theory and $SU(N)$ $\cN=4$ SYM -- arise in M-theory and string theory on the worldvolumes of stacks of $N$ M5 and D3 branes, respectively. Their partially twisted relatives are also realized by branes wrapping supersymmetric cycles in special holonomy manifolds \cite{Bershadsky:1995qy}. The explicit construction of the field theories in terms of wrapped branes is useful because there is a direct translation from the brane-geometric constructions of a field theory (which should be thought of as specifying its UV behavior) to boundary conditions for the dual supergravity solution.

As we are interested in the field theory limit of the brane dynamics, we should imagine the relevant supersymmetric cycles occurring in some compact, special holonomy manifold at large volume. In the large volume limit, the branes only probe an infinitesimal neighborhood of the supersymmetric cycle, so the geometry can be modeled as a non-compact manifold which is a vector bundle over $\cC$, where the fiber is $\RR^5$ in the case of M5 branes and $\RR^6$ in the case of D3 branes.\footnote{It is not necessary for the total space of this vector bundle to have a Ricci-flat metric, but only that such a metric exists in a neighborhood of the zero section of the vector bundle. This is because in the low energy limit, the tension of the branes effectively becomes infinite.} These fibers are precisely the vector spaces which appeared previously in Section \ref{ssec:twist} representing the field-theoretic R-symmetry groups. 

Accordingly, in the case of the $1/2$ BPS twist of both M5 and D3 theories, only a one-complex-dimensional subspace of the transverse space is fibered non-trivially over $\cC$. This amounts to the statement that $\cC$ is a holomorphic curve in a local Calabi-Yau two-fold of the form
\be
X_{1/2}=\cL\to\cC~,\label{2fold}
\ee
where $\cL$ represents a holomorphic line bundle. The condition that the R-symmetry component of the twisted rotation group acts on the preserved supercharges with equal and opposite charge to the untwisted rotation group specifies that this line bundle is in fact the holomorphic cotangent bundle $T^{\star}\cC^{(1,0)}$.\footnote{The choice of holomorphic, as opposed to anti-holomorphic, cotangent bundle is merely a convention.} This is the unique line bundle $\cL$ which admits a hyper-K\"ahler metric, and so leads to a theory with $\cN=2$ supersymmetry.

In the case of the $1/4$ BPS twists, there is a non-trivial $\CC^2$ bundle over $\cC$, and the twisted rotation group acts distinctly on the two $\CC$-factors. This situation arises when $\cC$ is a holomorphic curve in a local Calabi-Yau three-fold of the form
\be
X_{1/4}=\cL_1\oplus\cL_2\to\cC~.\label{threefold}
\ee
As mentioned in Section \ref{ssec:twist}, a variety of choices can be made for the line bundles $\cL_1$ and $\cL_2$ so that the resulting geometry is locally Calabi-Yau (which in turn ensures that supersymmetry on the branes is preserved).\footnote{The holomorphic structure on the $\CC^2$ bundle does not have to factorize in general, so there are geometries which are not sums of holomorphic line bundles. In the $1/4$ BPS twisted theory studied here, the holomorphic structure can be deformed to an unfactorized one by turning on $SU(2)$ Wilson lines on $\cC$ -- see \cite{Benini:2009mz}. The story for more general twists preserving four supercharges is currently under investigation \cite{inprogresstwists}.} We focus on the case where the R-symmetry factor of the twisted rotation group acts identically on the two line bundles, with half the weight of the action of the ordinary rotation group. In short, we set $\cL_1=\cL_2\equiv\cL_{1/4}$ with $\cL_{1/4}^{\otimes 2}=T^\star\cC^{(1,0)}$.\footnote{There are, of course, $2g$ different choices for $\cL_{1/4}$ which satisfy this condition. However, since we work on the covering space of $\cC$ and performing a quotient without additional action on sections of these line bundles, we choose the spin structure corresponding to periodic boundary conditions. We thank Eva Silverstein for pointing out this ambiguity.}

\subsection{Supergravity Ans\"atze}\label{ssec:ansatze}

We are studying theories whose microscopic behavior is controlled by maximally supersymmetric theories with well-known supergravity duals. Consequently, it is straightforward to fix the asymptotic form of the dual supergravity backgrounds. Here we outline precisely the Ans\"atze which provide the starting point for our calculations. We first describe the backgrounds dual to the twisted M5 brane theories. For the twisted D3 brane theories the procedure is analogous and is described succinctly.

\subsubsection{M5 brane Ans\"atze}\label{sssec:M5ansatz}

The $A_{N-1}$ $(2,0)$ theory is dual at large $N$ to eleven-dimensional supergravity in an $AdS_7\times S^4$ background, where the $S^4$ factor can be thought of as the boundary of the transverse $\RR^5$ to a stack of $N$ M5 branes. From the brane construction of the partially twisted $(2,0)$ theory, we see that the large $N$ dual should be an eleven-dimensional supergravity background which is asymptotically locally $AdS_7\times S^4$, but for which the topology at fixed value of the radial coordinate is an $S^4$ fibration over $\RR^{1,3}\times\cC$. The $S^4$ fibration at the boundary is determined by the $\RR^5$ fibration in the brane construction (\ie, the complex structure of the noncompact Calabi-Yau). Fortunately, there is a consistent truncation of eleven-dimensional supergravity on $S^4$ to the lowest Kaluza-Klein modes on the $S^4$ given by the maximal gauged supergravity in seven dimensions \cite{Nastase:1999cb,Nastase:1999kf}. Since the boundary conditions involve only the lowest Kaluza-Klein modes, the existence of the consistent truncation guarantees that we can work entirely in the language of the lower-dimensional gauged supergravity, and that all of the solutions we obtain can be uplifted to solutions of eleven-dimensional supergravity using explicit formulae from \cite{Nastase:1999cb,Nastase:1999kf} (see also \cite{Cvetic:1999xp}).

The maximal gauged supergravity in seven dimensions has an ordinary $SO(5)$ gauge group (dual to the R-symmetry) and an $SO(5)_c$ composite gauge group \cite{Pernici:1984xx}. The field content includes the metric, the $SO(5)$ gauge field, fourteen scalars parametrizing the coset $SL(5,\RR)/SO(5)_c$ and five three-form potentials transforming in the $\bf{5}$ of the $SO(5)$ gauge group. There are also four spin-3/2 fields and sixteen spin-1/2 fields transforming in the $\bf{4}$ and $\bf{16}$ of $SO(5)_c$, respectively. The complete action and supersymmetry variations of this theory were derived in \cite{Pernici:1984xx}. The bulk fields which are needed to match the partial twists of the $(2,0)$ theory at the boundary lie in a simple truncation of this theory to the metric, two Abelian gauge fields in the Cartan of the $SO(5)$ gauge group (encoding the fibration of the $S^4$, which has a reduced $U(1)\times U(1)$ structure), and two scalars which parameterize squashing deformations of the four-sphere.\footnote{There is also a three-form gauge potential in this truncation, but it vanishes identically for all solutions discussed in the present work.} This is precisely the truncation of \cite{Liu:1999ai}, but note that it is not
the bosonic part of a non-maximal supergravity. However it has been shown that every solution of the equations of motion of the truncated theory solves the equations of motion of the maximal theory \cite{Liu:1999ai,Cvetic:1999xp}.

It is now straightforward to write down the most general Ansatz appropriate to our construction. The seven-dimensional metric takes the form
\be
ds^2 = e^{2f} (-dt^2+dz_1^2+dz_2^2+dz_3^2) + e^{2h}dr^2 + y^{-2}e^{2g} (dx^2+dy^2)~. \label{M5metricAnsatz}
\ee
where $f$, $g$, and $h$ are functions of $r$ and of the coordinates $(x,y)$, which take values on the upper half-plane $H=\{(x,y)\,|\,y>0\}$.\footnote{For appropriate choices of the function $g$ and the range of $(x,y)$, this Ansatz is compatible with the Riemann surface $\cC$ having low genus (${\bf g}=0,1)$. Indeed, the derivations found in Appendix \ref{app:derivations} are sufficient to describe these cases.} In order to obtain a compact Riemann surface parameterized by $(x,y)$, we impose a quotient by a discrete (Fuchsian) subgroup $\Gamma\subset PSL(2,\RR)$, the automorphism group of the hyperbolic plane. The functions $f$, $g$, and $h$ must be invariant under $\Gamma$. In addition to the metric, there may be non-trivial $(r,x,y)$-dependent profiles for the two Abelian gauge fields and two real scalars in the truncation,
\be
 A^{(i)}= A^{(i)}_x dx +A^{(i)}_y dy+A^{(i)}_r dr~, \qquad\qquad \lambda_i=\lambda_i(x,y,r)~, ~~i=1,2~. \label{M5gaugescalarAnsatz}
\ee
These bosonic fields must also to transform covariantly under $\Gamma$.

As mentioned above, the asymptotic form of this Ansatz is fixed by the brane construction of the boundary theory. Specifically, the metric functions should have the following UV behavior as $r\to0$,
\begin{align}
\label{M5MetricAsymp}
\begin{split}
f(x,y,r), ~h(x,y,r) &\to-\log r+\cdots~,\\[0pt]
g(x,y,r) &\to-\log r + g_0(x,y)+\cdots~,
\end{split}
\end{align}
where $\cdots$ represents terms which vanish as $r\to0$. The asymptotic behavior of the bosonic fields is given by
\bea
\lambda_i&\to&0+\cdots~,\notag\\
A^{(i)}_r&\to&0+\cdots~, \label{M5ScalarAsymp}\\
A^{(i)}_{x,y}&\to&a^{(i)}\omega^{xy}_{x,y}+\cdots~,\notag
\eea
where $\omega_\mu$ is the spin connection in seven dimensions. The constants $a^{(i)}$ are determined by the choice of twist, and the condition \eqref{M5ScalarAsymp} for the gauge fields $A^{(i)}_{x,y}$ encodes the fact that at the boundary the $S^4$ fibration is completely specified by the structure of the tangent bundle to $\cC$. To be precise, in the $1/2$ BPS twist, the correct choice is $a^{(1)}=1/2m$, $a^{(2)}=0$, while for the $1/4$ BPS twist we take $a^{(1)}=a^{(2)}=1/4m$, where $m$ is the gauge coupling of the gauged supergravity.\footnote{The appearance of the parameter $m$ may look strange, since one might expect these values to match those of the parameters $a$ and $b$ which appeared in the discussion of Section \ref{sssec:M5twist}. This is a consequence of the standard normalization for gauge fields in gauged supergravity which differs by a factor of $2m$ from the more geometric normalization in which the gauge fields can be naturally interpreted as connections on principle bundles.} 

Moreover, the twists in question preserve additional symmetries which lead to simplifications for the bosonic scalar fields. In the case of the $1/2$ BPS twist, there is an $SU(2)$ global symmetry coming from the fact that the transverse $\RR^5$ has an $\RR^3$ factor which is fibered trivially. This leads to the simplification
\be
2\lambda_1+3\lambda_2=0~,\quad\quad A^{(2)}=0~,
\ee
which can be consistently imposed as a truncation at the level of the equations of motion. In the $1/4$ BPS twist, there is an extra $\ZZ_2$symmetry which exchanges $\cL_1$ and $\cL_2$ in the geometry \eqref{threefold}. This implies the additional relation
\be
\lambda_1=\lambda_2~,\quad\quad A^{(1)}=A^{(2)}~,
\ee
which again leads to a consistent truncation of the equations of motion.

\subsubsection{D3 brane Ans\"atze}\label{sssec:D3ansatz}

For the twisted D3 brane backgrounds, we have a very similar story. At large $N$ and large 't Hooft coupling, $\cN=4$ SYM with $SU(N)$ gauge group is dual to type IIB supergravity in $AdS_5\times S^5$, with the $S^5$ thought of as the boundary of the transverse $\RR^6$ to a stack of $N$ D3 branes. We expect the twisted theory to be dual to a background which is asymptotically locally $AdS_5\times S^5$ with the spacetime topology at fixed value of the radial coordinate given by an $S^5$ fibration over $\RR^{1,1}\times\cC$. The asymptotic $S^5$ fibration is determined by the $\RR^6$ fibration in the brane construction. 

It is again sufficient to work in a gauged supergravity description. The maximal gauged supergravity in five dimensions was constructed in \cite{Gunaydin:1984qu, Pernici:1985ju, Gunaydin:1985cu} where the full action and supersymmetry variations were derived, and it is believed to be a consistent truncation to the lowest Kaluza-Klein modes of type IIB supergravity on $S^5$.  This has not been proven explicitly, but in the present work we do not need the full structure of the theory. Rather, we content ourselves to work with the subsector studied in \cite{Cvetic:1999xp}. This is a truncation of the maximal theory to the metric, three Abelian gauge fields in the Cartan of the $SO(6)$ gauge group, and two real, neutral scalars. It can be shown to be a consistent truncation of the maximally supersymmetric supergravity to the bosonic part of an $\cN=2$ gauged supergravity coupled to two vector multiplets (see \cite{Bobev:2010de} for a recent discussion of this truncation). For this truncation, it {\it has} been shown that all solutions can be uplifted to solutions of type IIB supergravity, and there exist explicit uplift formulae \cite{Cvetic:1999xp}. Thus, all of the solutions discussed in the present work can be written as explicit solutions of type IIB supergravity.

The Ansatz for the twisted D3 brane solutions takes a form analogous to that of the twisted M5 solutions. The five-dimensional metric is
\be
ds^2 = e^{2f} (-dt^2+dz^2) + e^{2h}dr^2 + y^{-2}e^{2g}(dx^2+dy^2)~, \label{D3metricAnsatz}
\ee
and there are now three Abelian gauge fields and two real scalars,
\begin{align}
\begin{split}
&A^{I}= A^{I}_x dx +A^{I}_y dy+A^{I}_r dr~,\quad\quad I=1,2,3~,\\[0pt]
&\phi_1(x,y,r)~,\quad\quad \phi_2(x,y,r)~.
\end{split}
\end{align}
All functions in this Ansatz depend on $(x,y,r)$ and two-dimensional Poincar\'e invariance is manifest.

The behavior at $r\to0$ is controlled by the corresponding twist of $\cN=4$ SYM. The metric functions have the following asymptotics,
\begin{align}
\label{D3MetricAsymp}
\begin{split}
f(x,y,r), ~h(x,y,r)&\to-\log r+\cdots~,\\[0pt]
g(x,y,r)&\to-\log r + g_0(x,y)+\cdots~,
\end{split}
\end{align}
while the bosonic fields obey
\bea
\phi_{1,2}&\to&0+\cdots~, \notag\\
A^{I}_r&\to&0+\cdots~,\label{D3scalarUV}\\
A^{I}_{x,y}&\to&a^{(I)}\omega^{xy}_{x,y}+\cdots~.\notag
\eea
In this gauged supergravity, the effective gauge coupling is set to one, and the values of the constants $a^{(I)}$ are those of the constants $a$, $b$, and $c$ which appeared in Section \ref{sssec:D3twist}. In particular, for the $1/2$ BPS twist we have $a^{(1)}=a^{(2)}=0$ and $a^{(3)}=1$, while for the $1/4$ BPS twist we take $a^{(1)}=a^{(2)}=1/2$ and $a^{(3)}=0$.

For these choices of twists the backgrounds enjoy additional global symmetries which imply further constraints on the bosonic fields. Specifically, the presence of a $\ZZ_2$ symmetry of the geometry which descends to the $\cN=2$ gauged supergravity implies a global relation
\be
\phi_2=0~,\quad\quad A^{1}=A^{2}~.
\ee
For the $1/2$ BPS twist, this implies $A^{1}=A^{2}=0$, while for the $1/4$ BPS twist it yields $A^{3}=0$. These are both consistent truncations from the $U(1)^3$ gauged supergravity to theories with only a single gauge field and scalar. We are now prepared to derive the conditions for the backgrounds just discussed to preserve the appropriate amount of supersymmetry.

\sect{Holographic Flows for Twisted M5 Branes}\label{sec:M5s}

Our goal is to derive flow equations which describe the supersymmetric evolution of the background fields of Section \ref{sssec:M5ansatz} as a function of the radial coordinate and to understand their late-time, or IR, behavior as a function of the boundary metric on $\cC$ (the function $g_0(x,y)$ in \eqref{M5MetricAsymp}). The flow equations are determined by the condition that the bosonic background be invariant under an appropriate number of supersymmetry transformations, {\it i.e}, by the condition that the variations of all fermionic fields vanish in the background. The relevant supersymmetry variations for the fermionic fields in the truncated maximally supersymmetric gauged supergravity are given by  \cite{Pernici:1984xx,Liu:1999ai}
\bea
\delta\psi_{\mu} &=& \left[ \nabla_{\mu} +m(A_{\mu}^{(1)}\Gamma^{12} + A_{\mu}^{(2)}\Gamma^{34}) + \tfrac{m}{4} e^{-4(\lambda_1+\lambda_2)}\gamma_{\mu} + \tfrac{1}{2} \gamma_{\mu}\gamma^{\nu} \partial_{\nu} (\lambda_1+\lambda_2) \right] \epsilon \notag\\[3pt]
 && + \tfrac{1}{2} \gamma^{\nu}\left( e^{-2\lambda_1}F_{\mu\nu}^{(1)}\Gamma^{12} + e^{-2\lambda_2}F_{\mu\nu}^{(2)}\Gamma^{34} \right) \epsilon~, \notag \\ [5pt]
\delta\chi^{(1)} &=& \left[ \tfrac{m}{4} (e^{2\lambda_1}- e^{-4 (\lambda_1+\lambda_2)}) - \tfrac{1}{4} \gamma^{\mu} \partial_{\mu} (3\lambda_1+2\lambda_2) - \tfrac{1}{8} \gamma^{\mu\nu} e^{-2\lambda_1}F_{\mu\nu}^{(1)}\Gamma^{12} \right] \epsilon ~, \label{gendil1M5}\\[5pt]
\delta\chi^{(2)} &=& \left[ \tfrac{m}{4}(e^{2\lambda_2}- e^{-4 (\lambda_1+\lambda_2)}) - \tfrac{1}{4} \gamma^{\mu}   \partial_{\mu} (2\lambda_1+3\lambda_2) -\tfrac{1}{8} \gamma^{\mu\nu} e^{-2\lambda_2}F_{\mu\nu}^{(2)}\Gamma^{34} \right] \epsilon ~. \notag
\eea
The parameter $m$ is proportional to the gauge coupling constant of the supergravity and is inversely proportional to the scale of $AdS_7$. The analysis of these BPS conditions is described in detail in Appendix \ref{app:derivations}. The results are remarkably simple for both choices of twist. The full solutions to the BPS constraints are encoded in the solution to a system of two coupled partial differential equations (PDEs) for the metric function $g$ and a linear combination of the scalar fields $\lambda_i$. We first discuss the resulting flows for the $1/2$ BPS twist.

\subsection{1/2 BPS flows}\label{ssec:halfBPSM5}

For this choice of twist, the Ansatz from Section \ref{sssec:M5ansatz} imposes the relation
\be
2\lambda_1+3\lambda_2=0~,\quad\quad A^{(2)}=0~,
\label{M5n=2Trunc}
\ee
and we work in terms of a reduced set of bosonic fields defined as
\be
\lambda\equiv\lambda_2~,\quad\quad A~\equiv~ A^{(1)}~.
\ee
Applying the conditions for unbroken supersymmetry as described in Appendix \ref{app:derivations}, we find that the supersymmetric background is determined by the solution to the following system of PDEs,
\begin{align}
\label{BPSimpEqn}
\begin{split}
&\partial_\rho\lambda=-\ds\tfrac{2m}{5}+\ds\tfrac{2m}{5}e^{-5\lambda}+\ds\tfrac{1}{5m}e^{\lambda-2g}\left(1+\Delta (g+2\lambda)\right) ~,\\[5pt]
&\partial_\rho g=\ds\tfrac{3m}{10}+\ds\tfrac{m}{5}e^{-5\lambda}-\ds\tfrac{2}{5m}e^{\lambda-2g}\left(1+\Delta (g+2\lambda)\right)~,
\end{split}
\end{align}
with $\Delta \equiv y^2(\partial_{x}^2+\partial_{y}^2)$. The radial variable $\rho$ is defined in \eqref{AAEq24}. These flow equations can be further simplified by defining\footnote{In fact, we would like to think of $\nonlin$ as the conformal factor for an auxiliary metric on $\cC$. While it does not describe an actual metric which appears in the supergravity setting, it is in some sense the ``right'' metric from the point of view of the flow equations.}
\be
\nonlin (\rho,x,y) \equiv 2 g(\rho,x,y) + 4 \lambda(\rho,x,y)~,
\ee
with respect to which equations \eqref{BPSimpEqn} can be rewritten as
\be
\partial_{\rho}^2 e^{\nonlin} +\Delta\nonlin +2 - m^2 e^{\nonlin} = 0~,\label{heaveng}
\ee
along with a condition for $\lambda$ as a simple function of $\nonlin$,
\be
e^{-5\lambda} = \tfrac{1}{2m} (m+ \partial_{\rho} \nonlin)~. \label{heavenl}
\ee
There are a couple of curiosities to be noted about equation \eqref{heaveng}. First off, in terms of $\Phi(\rho,x,y) = \nonlin(\rho,x,y)- 2\log y$, the equation becomes
\be
(\partial_x^2+\partial_y^2)\Phi +  \partial_{\rho}^2 e^{\Phi} = m^2 e^{\Phi}~.
\label{M5N=2Heaven}
\ee
For $m=0$ this is the continuum $SU(\infty)$ Toda equation (also known as the Heavenly Equation, or Plebanski's Heavenly Equation). It is integrable and has been extensively studied (see, \eg, \cite{Boyer:1982,Bakas:1989,Saveliev:1993}). Since the parameter $m$ is inversely proportional to the scale of $AdS_7$, we necessarily have $m\neq 0$. We do not know whether the equation with $m\neq 0$ inherits any nice properties from the $m=0$ case. The $SU(\infty)$ Toda equation also appears in the analysis of \cite{Lin:2004nb} and \cite{Gaiotto:2009gz}, where the role of the variable $\rho$ is played by one of the coordinates on the topological four-sphere in the eleven-dimensional solution.
In addition,  \eqref{M5N=2Heaven} is time-reversal ($\rho$-reversal) invariant. This will not be the case for the other flows that we derive, and 
we do not know the repercussions of this symmetry.

In the remainder of this section, we perform a concrete analysis of the local properties of solutions to \eqref{heaveng}. We study the linearized behavior of solutions in the IR and UV, and also perform a perturbative analysis of solutions which are globally very close to the exact solutions of \cite{Maldacena:2000mw}. The analysis paints a picture where solutions behave as uniformizing flows for the metric on $\cC$ {\it locally} around the constant curvature metric. However, we find that the question of global behavior is intractable using direct methods. Section \ref{sec:theproof} contains a more abstract analysis of the global space of solutions, culminating in a proof that the flow equations we have derived are globally uniformizing.

\subsubsection{Infrared analysis}\label{sssec:halfBPSM5IR}

To begin, we determine the structure of four-dimensional conformal fixed points in the IR. Such a conformal point should be described by a supergravity background of the form $AdS_5\times\cC$, so in particular $\nonlin(\rho,x,y)$ should be constant with respect to $\rho$, and we are looking for fixed points of \eqref{heaveng} and \eqref{heavenl}. A fixed point of \eqref{heaveng} satisfies
\be
e^{-\nonlin}(2+\Delta\nonlin)-m^2=0~.\label{IREq1}
\ee
This is the Liouville equation for the function $\Phi/2$, which makes it clear that the only solution is
\be
e^{\nonlin_\sir}=\frac{2}{m^2}~.\label{IREq2}
\ee
Combining this with \eqref{heavenl} (and \eqref{AAEq11}--\eqref{AAEq18}) yields the fixed point values for all the background functions,
\be
e^{g}=\frac{2^{1/10}}{m}~,\quad\quad e^{\lambda}=2^{1/5}~,\quad\quad e^{f}=e^{h}=\frac{2^{3/5}}{m}\ds\frac{1}{r}~.\label{IREq3}
\ee
We conclude that even when the metric on $\cC$ is allowed to vary arbitrarily, the only $\cN=2$ $AdS_5$ vacua are those studied in \cite{Maldacena:2000mw}, for which the metric has constant negative curvature.

We can study the perturbative behavior of these solutions around the IR fixed point.\footnote{By virtue of \eqref{AAEq22} and \eqref{AAEq24}, the IR ($r\to\infty$) corresponds to $\rho\to-\infty$, and the UV ($r\to0$) corresponds to $\rho\to +\infty$.} This tells us about the late-time behavior of solutions which flow to the conformal fixed point \eqref{IREq3}. In particular, we anticipate that there should be linearized solutions in the IR for which the conformal factor is approaching its fixed point value from arbitrary directions in the space of metrics on $\cC$. 

We work with \eqref{heaveng} and study the expansion
\be
\nonlin = \nonlin_\sir + \varepsilon \lin(\rho,x,y)~,
\ee
to leading order in the infinitesimal parameter $\varepsilon$. We do not explicitly unpack our solutions in terms of the function $f$, $g$, $h$, and $\lambda$, but instead limit our discussion to $\nonlin$, which can be treated as a proxy for the behavior of the metric function $g$ and the scalar $\lambda$. To linear order in $\varepsilon$, $\lin(\rho,x,y)$ solves
\be
\partial_\rho^2\lin+ \tfrac{m^2}{2}~\Delta \lin -m^2 \lin = 0 ~.
\label{IRlinearPsi}
\ee
This is a linear PDE which we can solve by expanding $\lin$ in eigenfunctions of the Laplacian on the Riemann surface
\be
\lin  = \ds\sum_{n=0}^\infty \lin_n(\rho) Y^{(n)}(x,y)~.
\label{IRHarmonics}
\ee
Since the Riemann surface is compact and hyperbolic, we have
\be
\Delta  Y^{(n)}(x,y) = - \mu_n Y^{(n)}(x,y) ~, \qquad\qquad \mu_0=0~, ~~~ \mu_n>0~, ~~~ n>1~. \label{Yndefinition}
\ee
Inserting the expansion \eqref{IRHarmonics} into equation \eqref{IRlinearPsi}, we find the most general solution,
\be
\lin_n(\rho) = a_{n} e^{\alpha_{n}^{(+)}m\rho} + b_{n} e^{\alpha_{n}^{(-)}m\rho}~,\label{IRExpand}
\ee
where
\be
\alpha_{n}^{(\pm)}= \pm \sqrt{1+\tfrac{1}{2}\mu_n }~,
\ee
and $a_n$ and $b_n$ are free coefficients. For the solution to be regular in the IR, all of the $b_n$ must vanish. This leaves infinitely many solutions which approach the $AdS_5$ fixed point in the IR but for which the metric on $\cC$ is perturbed in the UV in an arbitrary way. This confirms our expectations that there should exist flows approaching the $AdS_5$ fixed point from all directions in the space of metrics on $\cC$, and we interpret all of these modes as irrelevant operators in the IR SCFT which may be turned on along the RG flow from six dimensions depending on the metric on $\cC$ in the UV. The modes with $b_n\neq0$, however, take the solution away from the $AdS_5$ fixed point in the IR. We expect these modes to generically be unphysical, with possible exceptions which we discuss briefly in Section \ref{sssec:halfBPSM5UV}. We conclude that in the neighborhood of the fixed point, the BPS flow equations exhibit an attractor type behavior in the space of metrics on $\cC$.

\subsubsection{Ultraviolet analysis}\label{sssec:halfBPSM5UV}

To perform a perturbative analysis in the UV, it is convenient to define a new radial variable $\zeta=e^{-\frac{m}{2}\rho}$. We can solve the system of coupled PDEs \eqref{BPSimpEqn} perturbatively for $\zeta \to 0$ and find
\begin{align}
\label{UVExpand}\begin{split}
& g(\rho,x,y) = - \log(\zeta) + g_0(x,y) +g_2(x,y) \zeta^2+ g_{4\ell}(x,y) \zeta^4\log\zeta +g_4(x,y) \zeta^4+ \cO(\zeta^5)~,\\[5pt]
& \lambda(\rho,x,y) =  \lambda_2(x,y) \zeta^2+ {\lambda}_{4\ell}(x,y) \zeta^4\log\zeta + \lambda_4(x,y) \zeta^4+ \cO(\zeta^5)~, 
\end{split}
\end{align}
where
\bea
&& \lambda_2(x,y) =\tfrac{1}{5m^2} e^{-2g_0(x,y)} (1+\Delta g_0(x,y))~, \qquad\ g_2(x,y) = 3 \lambda_2(x,y)~, \notag\\[5pt]
&& \lambda_{4\ell}(x,y) = - \tfrac{2}{5m^2}e^{-2g_0(x,y)}\Delta (g_2(x,y)+2\lambda_{2}(x,y))~, \qquad g_{4\ell}(x,y) = \tfrac{1}{2} {\lambda}_{4\ell}(x,y)~, \\[5pt]
&& g_{4}(x,y) = \tfrac{1}{2} \lambda_4(x,y) - \tfrac{1}{4m^4} e^{-4g_0(x,y)} (1+\Delta g_0(x,y))^2 + \tfrac{1}{4m^2} e^{-2g_0(x,y)} \Delta (g_2(x,y)+2\lambda_{2}(x,y))~.\notag
\eea
The functions $g_{0}(x,y)$ and $\lambda_4(x,y)$ are undetermined and represent the two functional degrees of freedom in the choice of boundary conditions for the second-order PDEs. The function $g_{0}(x,y)$ is the metric on the Riemann surface in the UV. 

To build some intuition about the meaning of the function $\lambda_4(x,y)$ it is useful to consider solutions of the form \eqref{UVExpand} which are independent of $x$ and $y$ -- \ie, those which were studied in \cite{Maldacena:2000mw}. A scalar $\phi$ in asymptotically locally $AdS_7$ space which is dual to an operator of dimension $D$ and which depends only on the radial variable has the following UV behavior
\be
\phi(\zeta) \sim \phi_{s} \zeta^{6-D} + \ldots + \phi_{v} \zeta^{D}+\ldots~,
\ee
where $\phi_s$ is related to the source and $\phi_v$ to the vev of the dual operator (see, \eg, \cite{Bianchi:2001kw}). We conclude that the scalar $\lambda$ is dual to an operator $\cO_{\lambda}$ of dimensions $D=4$ in the $(2,0)$ CFT, and for solutions with no dependence on the coordinates $(x,y)$, there is a source for $\cO_{\lambda}$ which is fixed by the curved geometry, whereas the vev for the operator appears as a free parameter. It is a well-known difference between holographic RG flows and Wilsonian RG that in the gravitational setting, one must specify both sources and vets in the UV to formulate an initial value problem. This introduces the complication that in general, an arbitrary choice of the vevs will be unphysical \cite{Gubser:2000nd}. Nevertheless, it was argued in \cite{Maldacena:2000mw} that these flows are indeed physical for any (constant) choice of $\lambda_4$, with the flow reaching the $AdS_5$ fixed point only if $\lambda_4=0$, and otherwise leading to a singular flow which was interpreted as being dual to either the Coulomb or Higgs phase of the field theory, depending on the sign.

In backgrounds for which the fields have non-trivial profiles on the boundary of $AdS_7$ the holographic dictionary is not straightforward, and we cannot offer precise statements about the field theory interpretation of the function $\lambda_4(x,y)$.\footnote{We thank Balt van Rees for numerous helpful discussions of the subtleties associated with the holographic dictionary in such cases.}  However, it stands to reason that for fixed $g_0(x,y)$, there exists a specific choice of $\lambda_4(x,y)$ which corresponds to the configuration where the branes are unperturbed in the transverse directions and so there is a flow to the $AdS_5$ fixed point. We then expect a one-dimensional family of values for $\lambda_4(x,y)$, generalizing the constant values in the $(x,y)$-independent case, which lead to flows representing non-zero, physical vevs for the operator $\cO_{\lambda}$. We expect flows for generic values of $g_0(x,y)$ and $\lambda_4(x,y)$ to be unphysical, as they would imply the existence of field theory vacua with arbitrary $(x,y)$-dependent expectation values for $\cO_\lambda$. It would be interesting to understand whether one could determine the physically admissible values of $\lambda_4(x,y)$ by imposing a criterion for allowable singularities such as that of \cite{Gubser:2000nd} or \cite{Maldacena:2000mw}.

\subsubsection{Exact solution and fluctuations}\label{sssec:halfBPSM5lin}

The discussions above demonstrate that supersymmetric flows exist for any boundary metric $g_0(x,y)$ in the UV, and that additionally supersymmetric flows exist which approach the constant curvature solution in the IR from all directions in the space of metrics on $\cC$. In this section we study explicit flows which interpolate between the two sets of asymptotics. The key fact that facilitates this analysis is that the flow equation \eqref{heaveng} admits an exact solution under the assumption that $\nonlin$ is a function of $\rho$ alone. This is the solution found by Maldacena and N\'u\~nez in \cite{Maldacena:2000mw}, which we denote by the subscript ``$\smn$'':
\be
e^{\nonlin_{\smn}} = \ds\frac{e^{2m\rho}+2 e^{m\rho}+C}{m^2e^{m\rho}}~.\label{MNSolution}
\ee
Here, $C$ is an integration constant which is proportional to the parameter $\lambda_{4}$ in \eqref{UVExpand} and represents the expectation value of the operator $\cO_{\lambda}$.\footnote{There is another integration constant parameterizing the freedom to shift $\rho$ by a constant amount. It is set to zero without loss of generality.} The flow ends at an $AdS_5$ fixed point only for $C=0$. For $C\neq 0$, one finds Coulomb/Higgs branch flows which diverge in the IR.

Thus we can study small perturbations of the exact solution \eqref{MNSolution} for {\it all} values of $\rho$. Such a perturbed solutions takes the form
\be
\nonlin(\rho,x,y) = \nonlin_{\smn}(\rho) + \varepsilon \lin(\rho,x,y)~.
\ee
The fluctuation term $\lin(\rho,x,y)$ can be expanded as
\be
 \lin(\rho,x,y) = \ds\sum_{n}  \lin_n(\rho) Y^{(n)}(x,y)~,
\ee
where $Y^{(n)}(x,y)$ are defined in \eqref{Yndefinition}. It is also convenient to define a new radial variable $\eta=e^{m\rho}$, where the IR now corresponds to $\eta \to 0$ and the UV to $\eta\to\infty$. With these definitions the linearization of \eqref{heaveng} for the functions $\lin_n(\rho)$ is
\bea
&& (\eta^3+2\eta^2+C\eta)  \ds\frac{d^2\lin_n}{d\eta^2} + (3\eta^2+2\eta-C)  \ds\frac{d\lin_n}{d\eta} - (2+\mu_n)\lin_n= 0~. \label{lneqn}
\eea
This equation admits an exact solution when $C=0$, \ie, when there is an $AdS_5$ fixed point in the IR.\footnote{There are exact solutions for other special values of $C$ but we do not study them since these flows are singular.} The solution can be written in terms of hypergeometric functions
\begin{multline}
\lin_n(\eta) =A_1^{(n)} \ds\frac{2^{\sigma_n}}{\eta^{\sigma_n}} \,_2F_1[-\sigma_n,2-\sigma_n;1-2\sigma_n; -\eta/2]\\
+A_2^{(n)} \ds\frac{\eta^{\sigma_n}}{2^{\sigma_n}} \,_2F_1[\sigma_n,2+\sigma_n;1+2\sigma_n; -\eta/2]~,
\label{solnPsin}
\end{multline}
where $A_1^{(n)}$ and $A_2^{(n)}$ are integration constants and we have defined 
\be
\sigma_n = \ds\sqrt{1+\tfrac{1}{2}\mu_n }\geq 1~.
\ee
The solutions with $A_1^{(n)}\neq 0$ are singular near $\eta =0$, so we set $A_1^{(n)}=0$. 

\begin{figure}[t]
\centering
\includegraphics[width=18cm]{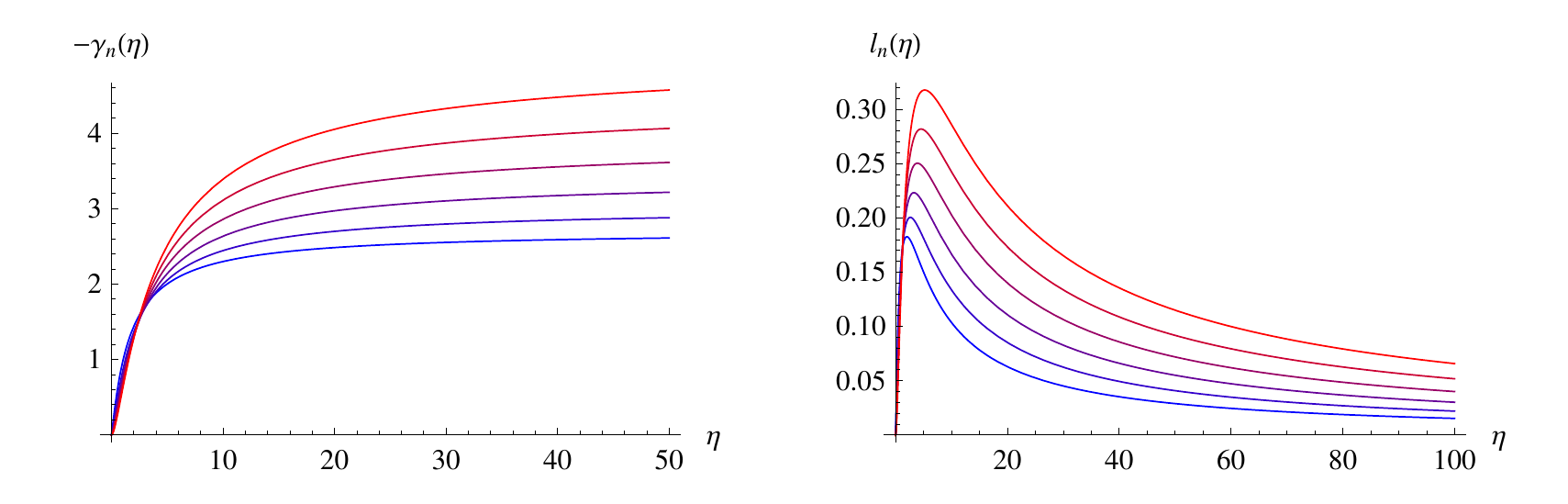}
\caption{{\it The functions $\gamma_n(\eta)$ (left) and $\ell_n(\eta)$ (right) for $B^{(n)}=1$ and $\mu_n=(1,2,3,4,5,6)$ (increasing as blue goes to red). The IR is at $\eta\to0$ and the UV is at $\eta\to\infty$.
    }}
\label{fig1}
\end{figure}

To get a better understanding of the physics of the linearized solution, it is helpful to write the functions $g$ and $\lambda$ as
\be
\lambda=\lambda_{\smn}(\rho)+\varepsilon\tilde{\lambda}(\rho,x,y) ~, \qquad\qquad g=g_{\smn}(\rho)+\varepsilon\tilde{g}(\rho,x,y)~,
\ee
and then expand $\tilde{\lambda}(\rho,x,y)$ and $\tilde{g}(\rho,x,y)$ in harmonics on the Riemann surface

\be
\tilde{\lambda} = \ds\sum_{n} \ell_n(\rho) Y^{(n)}(x,y)~, \qquad\qquad \tilde{g} = \ds\sum_{n} \gamma_n(\rho) Y^{(n)}(x,y)~.
\ee
The solutions for $\ell_n(\eta)$ and $\gamma_n(\eta)$ can be obtained from the exact solution for $\lin_n(\eta)$. The result, after applying standard identities for hypergeometric functions, is
\bea
 \ell_n &=&B^{(n)} \ds\frac{\eta^{\sigma_n}}{\eta^2+3\eta+2} \,_2F_1[\sigma_n,\sigma_n-1;2\sigma_n +1; -\eta/2]~,\\[5pt]
 \gamma_n &=& B^{(n)} \ds\frac{5 (\sigma_n-1)}{4(2\sigma_n^2+\sigma_n)}\ds\frac{\eta^{1+\sigma_n}}{2+\eta} \,_2F_1[\sigma_n+1,\sigma_n;2\sigma_n +2; -\eta/2]\notag \\[5pt]
 &&\qquad\qquad- B^{(n)} \ds\frac{4\sigma_n + 5(1+\eta)}{2\sigma_n(2+\eta)(1+\eta)} \,\eta^{\sigma_n} \,_2F_1[\sigma_n,\sigma_n-1;2\sigma_n +1; -\eta/2] ~,
\eea
where the new integration constants $B^{(n)}$ are proportional to $A_2^{(n)}$. The solutions for $\ell_n$ and $\gamma_n$ with $B^{(n)}=1$ and a range of values for $\mu_n$ are plotted in Figure \ref{fig1}. The expansion of $\ell_n$ in the UV ($\eta \to \infty$) and IR ($\eta \to 0$) is
\begin{align}
\begin{split}
 \ell_n|_{\eta\to\infty} &\approx ~B^{(n)} \left(\ds\frac{2^{\sigma_n-1} \Gamma(2\sigma_n+1)}{\Gamma(\sigma_n) \Gamma(\sigma_n+2)}~ \ds\frac{1}{\eta} +\cO(\log(\eta)\eta^{-2})\right)~,\\[5pt]
\ell_n|_{\eta \to 0} &\approx ~ B^{(n)} \left(\ds\frac{1}{2} \eta^{\sigma_n} - \left(\ds\frac{3}{4} +\ds\frac{\sigma_n(\sigma_n-1)}{4(2\sigma_n+1)}\right) \eta^{\sigma_n+1} +\cO(\eta^{\sigma_n+2})\right)~. 
\end{split}
\end{align}
Similarly, the UV/IR expansions of $\gamma_n$ are
\begin{align}
\begin{split}
 \gamma_n|_{\eta \to \infty} &\approx ~ -\ds\frac{5\,B^{(n)}\,2^{\sigma_n-2}}{\sigma_n^2} \left(\ds\frac{\Gamma(2\sigma_n+1)}{\Gamma(\sigma_n)\Gamma(\sigma_n+2)} +\cO(\eta^{-1})\right)~,\\[5pt]
\gamma_n|_{\eta \to 0} &\approx ~ B^{(n)} \left(1+\ds\frac{5}{4\sigma_n}\right) \eta^{\sigma_n}+\cO(\eta^{\sigma_n+1})~. 
\end{split}
\end{align}
Thus, these are interpolating solutions which fit the asymptotic expansions \eqref{IRExpand} and \eqref{UVExpand} with matching conditions
\begin{align}
\label{matchingconditions}
\begin{split}
a_{n}&=B^{(n)}\left(4+\frac{5}{2\sigma_n}\right)~,\\[5pt]
g_0(x,y)&=-5\sum_{n=0}^{\infty}B^{(n)}\frac{2^{\sigma_n-2}\Gamma(2\sigma_n+1)}{\sigma_n^2\Gamma(\sigma_n)\Gamma(\sigma_n+2)}Y^{(n)}(x,y)~.
\end{split}
\end{align}
The coefficients $B^{(n)}$ parameterize a neighborhood of the constant conformal factor for the boundary metric on $\cC$, and these interpolating flows demonstrate the conjectured uniformizing behavior for the metric in this neighborhood.  

\subsection{1/4 BPS flows}\label{ssec:quartBPSM5}

Turning to the $1/4$ BPS twist, the appropriate truncation of the seven-dimensional supergravity fields from Section \ref{sssec:M5ansatz} is
\be
A~\equiv~ A^{(1)}=A^{(2)}~, \qquad\qquad \phi\equiv-2\lambda_1=-2\lambda_2~.
\ee
The conditions for backgrounds respecting this truncation to preserve one quarter of the maximal supersymmetry are derived in Appendix \ref{app:derivations}. When these conditions are reformulated as flow equations intrinsic to the Riemann surface $\cC$, the resulting PDEs are
\begin{align}
 \label{BPSM5N1Eqn}
\begin{split}
&\partial_\rho\phi=- \tfrac{2m}{5} + \tfrac{2m}{5}e^{-5\phi} +\tfrac{1}{10m}e^{-3\phi-2g}\left(1+\Delta (g+4\phi)\right),\\[5pt]
&\partial_\rho g=\tfrac{m}{10} + \tfrac{2m}{5}e^{-5\phi} - \tfrac{2}{5m}e^{-3\phi-2g}\left(1+\Delta (g+4\phi)\right), 
\end{split}
\end{align}
with the new radial variable $\rho$ defined in \eqref{M5n=1Radial2}. As a second-order PDE for
\be
\nonlin(\rho,x,y) \equiv 2 g(\rho,x,y) + 8\phi(\rho,x,y)~,
\ee
these flow equations assume the form
\be
\Delta\nonlin + \partial_{\rho}^2 e^{\nonlin} -e^{\nonlin}\left( \tfrac{1}{2}(\partial_{\rho}\nonlin)^2 -m \partial_{\rho}\nonlin +\tfrac{3m^2}{2}\right) +2 = 0~.
\label{M51/4Heaven}
\ee
The scalar field $\phi$ is determined by $\nonlin(\rho,x,y)$,
\be
e^{-5\phi}=\tfrac{1}{4m}\left(3m+\partial_\rho\nonlin\right)~.
\ee
It is notable that while the first-order equations \eqref{BPSM5N1Eqn} are schematically similar to \eqref{BPSimpEqn}, the second-order equation \eqref{heaveng} appears much simpler than its $1/4$ BPS analogue \eqref{M51/4Heaven}. This foreshadows our inability to find any analytic  solutions of \eqref{M51/4Heaven}. Nevertheless, we are still able to perform a global analysis of solutions to \eqref{M51/4Heaven} in Section \ref{sec:theproof}.

\subsubsection{Infrared analysis}\label{sssec:quartBPSM5IR}

The unique $AdS_5$ vacuum in this truncation is determined by the constant solution of \eqref{M51/4Heaven},
\be
e^{\nonlin_\sir}= \ds\frac{4}{3m^2}~.
\ee
The background fields then take fixed point values
\be
e^{g}=\left(\ds\frac{3}{4}\right)^{\frac{3}{10}} \ds\frac{1}{m}~,\quad\quad e^{\phi}= \left(\ds\frac{4}{3}\right)^{\frac{1}{5}}~,\quad\quad e^{f}=e^{h}= \ds\frac{3^{\frac{4}{5}}}{2^{\frac{3}{5}} m} \ds\frac{1}{r}~.
\label{M5N1IRvalues}
\ee
We consider the following infinitesimal perturbation around the IR fixed point,
\be
\nonlin = \nonlin_\sir + \varepsilon \lin(\rho,x,y)~.
\ee
To leading order in $\varepsilon$, the perturbation $\lin(\rho,x,y)$ then obeys
\bea
&&\partial_{\rho}^2\lin + m\partial_{\rho}^2\lin -\tfrac{3m^2}{2}~\lin  + \tfrac{3m^2}{4}~\Delta  (\lin) =0 ~. \label{IRM5N1linearPsi}
\eea
By expanding $\lin(\rho,x,y)$ in harmonics on the Riemann surface as in \eqref{IRHarmonics}, we find the following solutions for $\lin_n(\rho)$,
\be
\lin_n(\rho) = a_{n} e^{\alpha_{n}^{(+)}\rho} + b_{n} e^{\alpha_{n}^{(-)}\rho}~,
\ee
where
\be
\alpha_{n}^{(\pm)}= -\tfrac{1}{2} \pm \tfrac{1}{2} \sqrt{7+3\mu_n}~.
\ee
Regularity of the solution in the IR requires that $b_n=0$ for all $n$.

\subsubsection{Ultraviolet analysis}\label{sssec:quartBPSM5UV}

Defining $\zeta=e^{-\frac{m}{2}\rho}$, the perturbative solution to equations \eqref{BPSM5N1Eqn} in the UV ($\zeta\to0$) is given by
\begin{align}
\begin{split}
& g(\rho,x,y) \approx - \log(\zeta) + g_0(x,y) +g_2(x,y) \zeta^2+ {g}_{4\ell}(x,y) \zeta^4\log\zeta +g_4(x,y) \zeta^4+ \cO(r^5)~,\\[5pt]
& \phi(\rho,x,y) \approx  \phi_2(x,y) \zeta^2+ {\phi}_{4\ell}(x,y) \zeta^4\log\zeta + \phi_4(x,y) \zeta^4+ \cO(r^5)~, 
\end{split}
\end{align}
where
\bea
&& \phi_2(x,y) =\tfrac{1}{10m^2} e^{-2g_0(x,y)} (1+\Delta g_0(x,y))~, \qquad\ g_2(x,y) = 6 \phi_2(x,y)~, \qquad {g}_{4\ell}(x,y) = {\phi}_{4\ell}(x,y)~, \notag\\[5pt]
&& {\phi}_{4\ell}(x,y) = \tfrac{1}{5m^4} e^{-4g_0(x,y)} (1+\Delta g_0(x,y))^2 - \tfrac{1}{5m^2} e^{-2g_0(x,y)} \Delta (g_2(x,y)+4\phi_{2}(x,y))~, \\[5pt]
&& g_{4}(x,y) =  \phi_4(x,y) - \tfrac{3}{8m^4} e^{-4g_0(x,y)} (1+\Delta g_0(x,y))^2 + \tfrac{1}{4m^2} e^{-2g_0(x,y)} \Delta (g_2(x,y)+4\phi_{2}(x,y))~.\notag
\eea
The undetermined function $g_{0}(x,y)$ is the conformal factor of the metric on $\cC$ and can be chosen arbitrarily. The function $\phi_4(x,y)$ is related to the vev of the dimension four operator $\cO_{\phi}$ dual to the supergravity scalar $\phi$. For fixed $g_0(x,y)$, we expect generic values of $\phi_4(x,y)$ to be unphysical, and for a unique value of $\phi_4(x,y)$ to lead to an $AdS_5$ vacuum in the IR.

\begin{figure}[!ht]
\begin{center}
\includegraphics[width=8.5cm]{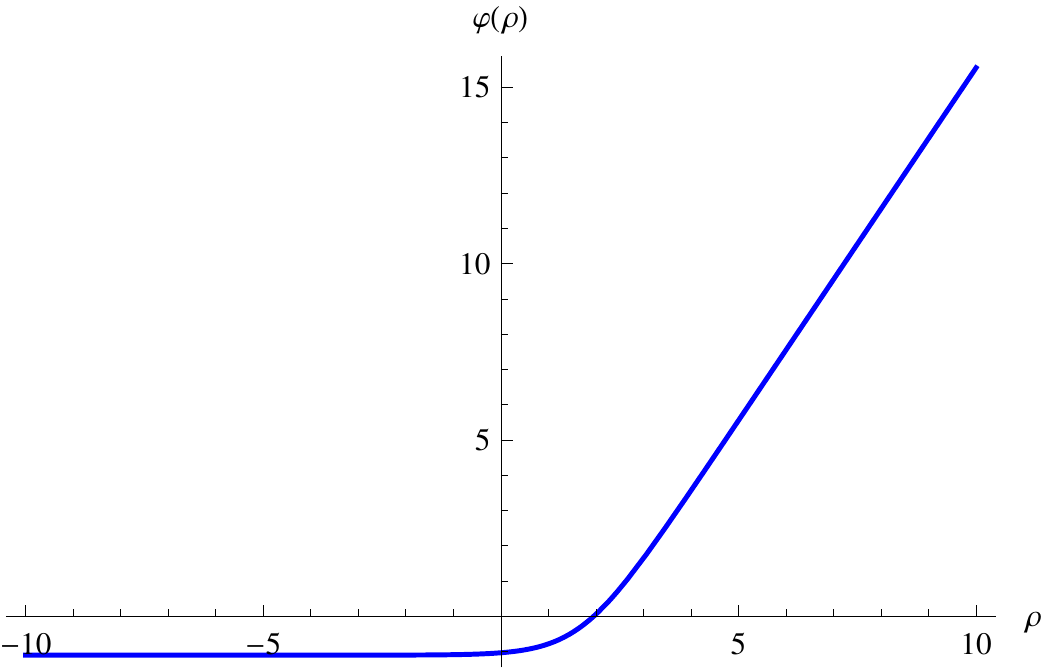}
\caption{\it A numerical solution for $\nonlin(\rho)$ for $1/4$ BPS M5 brane backgrounds. In the IR ($\rho\to-\infty$), the function approaches a constant, while in the UV ($\rho\to +\infty$), it diverges linearly (logarithmically in $r$).}
\label{M5numericalfigure}
\end{center}
\end{figure}

The absence of an analytic, constant-curvature flow equivalent to \eqref{MNSolution} in this case inhibits a direct study of the uniformizing behavior of these flows. However, when the background fields depend only on $\rho$, equations \eqref{BPSM5N1Eqn} do admit numerical solutions for which $g(\rho)$ and $\phi(\rho)$ have the prescribed asymptotic behavior of \eqref{M5N1IRvalues} and \eqref{M5MetricAsymp}--\eqref{M5ScalarAsymp}. A numerical solution is presented in Figure \ref{M5numericalfigure}. The existence of such a solution is important for the discussion in Section \ref{sec:theproof}.

\sect{Holographic Flows for Twisted D3 Branes}\label{sec:D3s}

The analysis of partially twisted D3 branes on $\cC$ closely parallels that of M5 branes. We look for flow equations which control the behavior of the background functions in the Ans\"atze of Section \ref{sssec:D3ansatz}. The problem is formulated in terms of the five-dimensional $\cN=2$ supergravity discussed in \cite{Cvetic:1999xp}, and the supersymmetry variations for the fermions are given in \cite{Behrndt:1998ns},
\begin{align}
\begin{split}
\delta\psi_{\mu} &= \left[ \nabla_{\mu} +\tfrac{i}{8} X_{I}(\gamma_{\mu}^{\nu\rho} - 4 \delta_{\mu}^{\nu} \gamma^{\rho}) F^{I}_{\nu\rho} + \tfrac{1}{2} X^{I}V_{I} \gamma_{\mu} - \tfrac{3i}{2} V_{I}A^{I}_{\mu} \right] \epsilon~, \\[5pt]
\delta\lambda_{(j)} &= \left[ \tfrac{3}{8} (\partial_{\phi_j} X_{I} )F^{I}_{\mu\nu} \gamma^{\mu\nu} + \tfrac{3i}{2} V_{I}\partial_{\phi_j} X^{I} - \tfrac{i}{4} \delta_{jk} \partial_{\mu}\phi_k \gamma^{\mu} \right] \epsilon~, \qquad j=1,2 ~,
\end{split}
\end{align}
where we define
\begin{align}
\begin{split}
& X^{1} \equiv e^{-\tfrac{1}{\sqrt{6}} \phi_1-\tfrac{1}{\sqrt{2}}\phi_2}~, \qquad\qquad X^{2} \equiv e^{-\tfrac{1}{\sqrt{6}} \phi_1+\tfrac{1}{\sqrt{2}}\phi_2}~, \qquad\qquad X^{3} \equiv e^{\tfrac{2}{\sqrt{6}}\phi_1}~, \\[5pt] 
& V_{I}=\tfrac{1}{3}~, \qquad\qquad\qquad\quad\,\, X_{I} = \tfrac{1}{3} (X^{I})^{-1}~.
\end{split}
\end{align}
Supersymmetric backgrounds of the $\cN=2$ theory preserve at most eight supercharges, and we study solutions which preserve only {\it two}, corresponding to $(1,1)$ supersymmetry in two dimensions. This is because the $\cN=2$ supergravity is a truncation of the maximally supersymmetric gauged supergravity for which the only visible supersymmetries are those generated by the spinor transforming with charges ${(\frac{1}{2},\frac{1}{2},\frac{1}{2})}$ under the $U(1)^3$ Cartan of $SO(6)_R$ (see \eqref{D3spinordecomp}). Both twists we study should preserve exactly two of these supercharges, but in the maximal gauged supergravity there are additional preserved supersymmetries which act identically on the fields in our truncation.

\subsection{$\cN=(4,4)$ flows}\label{ssec:halfBPSD3}

To find a BPS flow that preserves half of the maximum supersymmetry (\ie, 8 real supercharges) one should set
\be
\phi_2=0~, \qquad \alpha \equiv \tfrac{1}{\sqrt{6}} \phi_1~, \qquad A^{(1)} = A^{(2)}=0~, \qquad  A~\equiv~ A^{(3)}~.
\ee
The system of coupled PDEs intrinsic to $\cC$ is derived in Appendix \ref{app:derivations} and is given by
\begin{align}
\label{BPSimpD3N2Eqn}
\begin{split}
&\partial_\rho\alpha= 2 - 2e^{3\alpha} - e^{-\alpha-2g} (1+\Delta(g - \alpha))~,\\[5pt]
&\partial_\rho g = 2 + e^{3\alpha} - e^{-\alpha-2g} (1+\Delta(g - \alpha))~. 
\end{split}
\end{align}
This system of equations can be rewritten as a single second-order PDE
\be
\partial_{\rho}^2e^{\nonlin} -6 \partial_{\rho}e^{\nonlin} + 9 \Delta\nonlin +18=0~,
\label{D3N=44Heaven}
\ee
where
\be
\nonlin(\rho,x,y) \equiv 2 g(\rho,x,y) - 2 \alpha(\rho,x,y) ~,
\ee
and the scalar $\alpha$ is determined according to
\be
e^{3\alpha} = \tfrac{1}{6} \partial_{\rho}\nonlin~.\label{D3N=44Scalar}
\ee
As mentioned in Section \ref{sssec:D3twist}, the 1/2 BPS twist of $\cN=4$ SYM flows to an IR CFT which is a sigma model onto the Hitchin moduli space $\cM^H(\cC)$. It was pointed out in \cite{Maldacena:2000mw} that because this is a non-compact target space, one does not expect a normalizable, conformally invariant ground state for the theory, \ie, there should be no $AdS_3$ region in the gravity solution. This is also manifest in \eqref{D3N=44Scalar} which does not admit a constant solution for $\nonlin$ with finite $\alpha$.

\subsection{$\cN=(2,2)$ flows}\label{ssec:quartBPSD3}

To find gravity backgrounds dual to the 1/4 BPS twist of $\cN=4$ SYM, one should set
\be
\phi_2=0~, \qquad \alpha \equiv \tfrac{1}{\sqrt{6}} \phi_1~,  \qquad A~\equiv~ A^{(1)} = A^{(2)}~, \qquad A^{(3)} = 0~.
\ee
The system of coupled PDEs intrinsic to $\cC$ is
\begin{align}
\label{BPSimpD3N1Eqn}
\begin{split}
&\partial_\rho\alpha= - 2 + 2e^{-3\alpha} +\tfrac{1}{2} e^{-\alpha-2g} (1+\Delta(g+2\alpha))~,\\[5pt]
&\partial_\rho g =1 + 2e^{-3\alpha} - e^{-\alpha-2g} (1+\Delta(g+2\alpha))~.
\end{split}
\end{align}
While the second-order PDE that governs the flow is
\be
\partial_{\rho}^2e^{\nonlin} -\tfrac{1}{2} e^{\nonlin} (\partial_{\rho}\nonlin)^2 + 9 \Delta\nonlin +18 - 18 e^{\nonlin}=0~,
\label{D3N=22Heaven}
\ee
where we have defined
\be
\nonlin(\rho,x,y) \equiv 2 g(\rho,x,y) + 4 \alpha(\rho,x,y) ~.
\ee
The scalar $\alpha$ is related to $\nonlin$ via
\be
e^{-3\alpha} = \tfrac{1}{12} (6+\partial_{\rho}\nonlin)~.
\ee
The global properties of this equation are studied in more detail in Section \ref{sec:theproof}. Using \eqref{BPSimpD3N1Eqn} one can show that the metric on the Riemann surface in the UV can be arbitrary. For the $(x,y)$-independent solution, the UV asymptotic analysis of the system of flow equations was performed in Appendix A of \cite{Maldacena:2000mw} and we do not repeat it here. It is important to note that this linearized UV analysis suggests that in the dual twisted theory there is an operator of dimension two that triggers the RG flow. 

\subsubsection{Infrared analysis}\label{sssec:quartBPSD3IR}

The constant solution of \eqref{D3N=22Heaven} is given by
\be
e^{\nonlin_\sir}= 1~,
\ee
which implies the existence of a unique $AdS_3$ vacuum with the following scalar and metric functions:
\be
e^{\alpha}= 2^{1/3}~,\quad\quad e^{g}=2^{-2/3}~,\quad\quad  e^{f}=e^{h}= 2^{-2/3} \ds\frac{1}{r}~.
\label{D3N1IRvalues}
\ee
To study the BPS flow equations perturbatively around this fixed point, we write
\be
\nonlin = \nonlin_{\sir} + \varepsilon \lin(\rho,x,y)~.
\ee
After expanding \eqref{D3N=22Heaven} to linear order in $\varepsilon$ one finds the following equation for $\lin$,
\bea
&&\partial_{\rho}^2\lin -18~\lin  + 9~\Delta  \lin=0~.\label{IRD3N1linearPsi}
\eea
This equation can be solved by expanding in harmonics on the Riemann surface as defined by \eqref{Yndefinition},
\be
\lin = \ds\sum_{n=0}^{\infty} \lin_n(\rho) Y^{(n)}(x,y)~.
\ee
Solving for $\lin_n(\rho)$ then yields
\be
\lin_n(\rho) = a_{n} e^{\beta_{n}^{(+)}\rho} + b_{n} e^{\beta_{n}^{(-)}\rho}~,
\ee
where
\be
\beta_{n}^{(\pm)}= \pm 3 \sqrt{2+\mu_n}~.
\ee
As is by now familiar, regularity of the solution requires that the coefficients $b_n$ vanish.

\begin{figure}[!ht]
\begin{center}
\includegraphics[width=8.5cm]{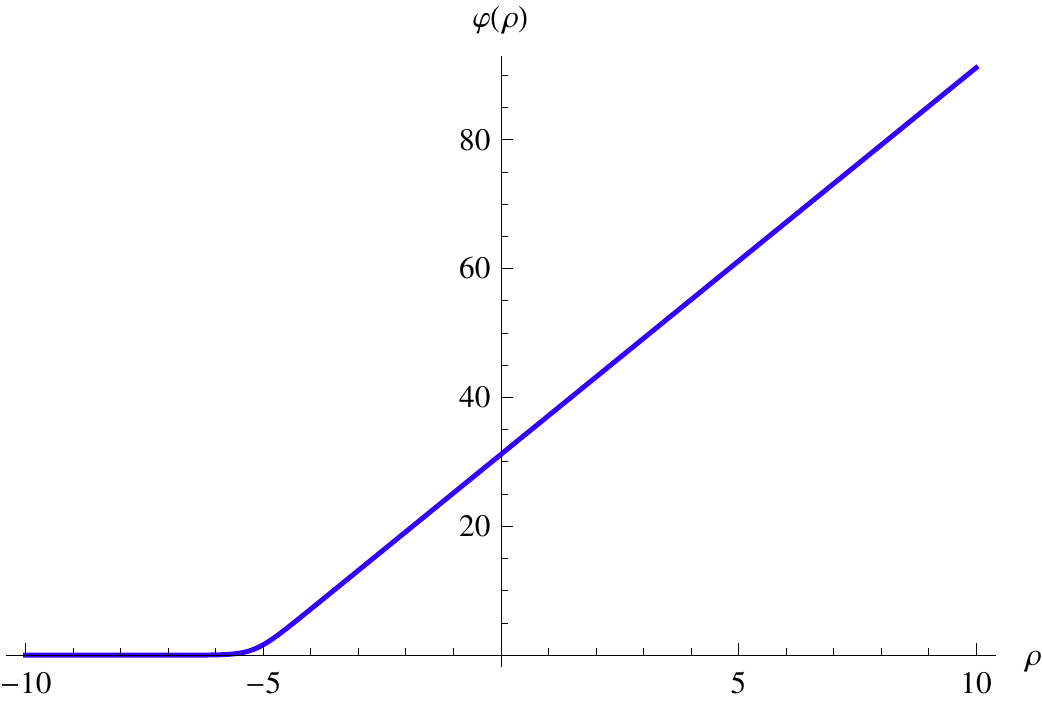}
\caption{\it A numerical solution for $\nonlin(\rho)$ for $\cN=(2,2)$ D3 brane solutions. In the IR ($\rho\to-\infty$), the function approaches a constant, while in the UV ($\rho\to +\infty$), it diverges linearly (logarithmically in $r$).}
\label{D3numericalfigure}
\end{center}
\end{figure}

We finally note that there are numerical solutions to equations \eqref{D3N=22Heaven} that depend only on $\rho$ (see Figure \ref{D3numericalfigure}). These solutions manifest the IR and UV behavior described by \eqref{D3N1IRvalues} and \eqref{D3MetricAsymp}--\eqref{D3scalarUV}, and their existence is important for the global analysis in the next section.

\sect{Global Analysis}\label{sec:theproof}

In this section, we prove that there exist solutions to equations \eqref{heaveng}, \eqref{M51/4Heaven}, and \eqref{D3N=22Heaven} for arbitrarily prescribed initial data in the UV which are asymptotic to the standard $(x,y)$-independent solution in the IR. We first describe in general terms a standard methodology for solving such problems, and then discuss the details for each specific case. 

To begin, let $M$ be a manifold with boundary.\footnote{The main example at hand is $M=\RR \times \cC$, so the boundary may be an asymptotic boundary.} Consider a real, scalar function $\nonlin\in C^{\infty}(M)$ and a nonlinear (elliptic) PDE in $\nonlin$,
\be
\PDO: C^{\infty}(M) \to C^{\infty}(M)~.
\ee
The basic issue is the solvability of the Dirichlet problem for $\PDO$, \ie, given boundary data $\dnonlin \in C^{\infty}(\dm)$, finding a scalar function $\nonlin$ such that
\be \label{5.1}
\PDO(\nonlin) = 0~, \quad\quad \nonlin|_{\dm} = \dnonlin~.
\ee
In the case of a boundary at infinity, the boundary value(s) must be understood asymptotically. Let $\cM$ be the on-shell moduli space of solutions $\nonlin$ to $\PDO(\nonlin) = 0$. The solvability of the Dirichlet problem above is equivalent to the surjectivity of the boundary map
\be \label{5.2}
\Bound: \cM \to C^{\infty}(\dm)~, \quad\quad \Bound(\nonlin) = \nonlin|_{\dm}~.
\ee

Before turning to the general approach for the infinite-dimensional setting, let us consider a toy model of finite-dimensional manifolds. Let $\pi: N_{1}^{d} \to N_{2}^{d}$ be a smooth map between compact $d$-manifolds (without boundary). A standard way to prove that $\pi$ is surjective is to calculate the (mod 2) degree $\deg(\pi) \in \ZZ_{2}$, defined as follows \cite{GP}. Let $q \in N_{2}$ be any regular value of $\pi$, \ie,~the derivative $D_{p}\pi: T_{p}N_{1} \to T_{q}N_{2}$ is surjective for any $p$ in the fiber $F_{q} = \pi^{-1}(q)$. The regular value property and compactness of $N_{1}$ imply that the cardinality $\# F_{q}$ is finite and $\deg(\pi) \equiv \# F_{q}$ (mod 2). That the degree is independent of the choice of regular value $q$ can be reasoned as follows. For $q_{1}\neq q_{2}$ both regular values of $\pi$, consider a generic path $\g \subset N_{2}$ which joins them. Then the inverse image $\pi^{-1}(\g) \subset N_{1}$ is a collection of one-manifolds -- paths or circles $\{\s_{i}\}$ with endpoints in the fibers $F_{q_{1}}$ and $F_{q_{2}}$. Those $\s_{i}$ which are open paths either join a point in $F_{q_{1}}$ with a point in $F_{q_{2}}$, or begin and end in a fixed fiber $F_{q_{i}}$. Since all points in the fibers are accounted for in this way, the cardinality mod 2 is independent of the choice of regular value $q$. If $\pi$ is not surjective, any point  $q \notin {\rm im}(\pi)$ is a regular value of $\pi$ (by definition) and $\# \pi^{-1}(q) = 0$. Thus it follows that if $\deg \, \pi \neq 0$, then $\pi$ is surjective. The concept of degree above can be extended to a $\ZZ$-valued degree given appropriate orientations, but we forgo such issues here. 

Under appropriate conditions, a similar methodology can be applied for infinite-dimensional (function) spaces. The $\ZZ_{2}$ degree is then known as the Smale degree \cite{Sm}, and is closely related to the Leray-Schauder degree. For general background in related topics of nonlinear functional analysis or global analysis, see \cite{N1,N2,E}.  The general procedure has the following three parts.

\medskip

\begin{description}
\item[I - Local Theory] {\ }

Prove that the on-shell moduli space $\cM$ is a smooth, infinite-dimensional (Banach or Hilbert) manifold, and that the boundary map $\Bound$ in \eqref{5.2} is a smooth Fredholm map with Fredholm index zero. The issue of whether $\cM$ is a manifold is equivalent to the issue of ``linearization stability'', \ie, at any solution $\nonlin$ of \eqref{5.1}, any solution $\lin$ to the linearized equation $D\PDO_{\nonlin}({\tilde\nonlin}) = 0$ is tangent to a curve $\nonlin_{t}$ of solutions of the nonlinear equation \eqref{5.1}. The usual method to prove that $\cM$ is a smooth manifold is to use the ``regular value theorem", (a version of the implicit function theorem): $\cM = \PDO^{-1}(0)$ is a manifold if $0$ is a regular value of $\PDO$, (the derivative $D\PDO$ is surjective at any point in $\cM$).

The Fredholm property means that the linearization, or derivative map, $D\Bound \equiv D_{\nonlin}\Bound$ at any point $\nonlin \in \cM$ has finite-dimensional kernel and cokernel, and the range of $D\Bound$ is closed. The Fredholm index is defined as 
\be
{\rm ind}(D\Bound)=\dim\left(\ker\left(D\Bound\right)\right) - \codim({\rm im}(D\Bound))~.
\ee
For example, self-adjoint operators have index zero. The requirement of Fredholm index zero essentially means that $D\Bound$ is an isomorphism modulo finite-dimensional factors of equal dimension. The manifold property of $\cM$ and the Fredholm property of $\Bound$ are closely related, and usually treated concurrently. These properties depend on choosing suitable function spaces (degrees of smoothness) in which to carry out the analysis; one typically chooses spaces in which there are good elliptic regularity properties. 
\medskip
  
\item[II - Compactness]{\ }

The moduli space $\cM$ and space of boundary data $C^{\infty}(\dm)$ are non-compact and infinite-dimensional. The key ingredient needed for the degree to continue to make sense is that the boundary map \eqref{5.2} is a proper map: if ${\cal K}$ is any compact set in the target $C^{\infty}(\dm)$, then the inverse image $\Bound^{-1}({\cal K})$ is a compact set in $\cM$. This means that for $\nonlin^{(i)}\in \cM$ a sequence of solutions with boundary data $\dnonlin^{(i)}$, convergence of the boundary data implies convergence of the solutions. In essence, this is the statement that the boundary data $\dnonlin$ controls the bulk solution $\nonlin$ with $\Bound(\nonlin) = \dnonlin$, and in practice, this amounts to proving ``{\it a priori} estimates" for $\nonlin$ in terms of $\dnonlin$. If $\Bound$ is proper, then $\# \Bound^{-1}(\dnonlin)$ is finite. 
\medskip

\item[III - Degree Calculation]{\ }

If the first two steps can be carried out, then the Smale degree $\deg\Bound \in \ZZ_{2}$ is well-defined. The argument in the toy model above then works in the same way for infinite-dimensional manifolds \cite{Sm}. To prove surjectivity of $\Bound$ in \eqref{5.2}, one then needs to show that 
\be \label{5.4}
{\rm deg}\,\Bound = 1~.
\ee
This is typically done by showing that there is a standard solution $\nonlin_{\textsc{ss}}$ -- \eg,~the $(x,y)$-independent solutions for the flows we consider here -- and then showing that this solution is the unique solution with its boundary data and that $\nonlinss$ is a regular point of $\Bound$. This establishes the surjectivity of $\Bound$.\footnote{Note that the process described here establishes global existence of solutions to \eqref{5.1}, but does not prove global uniqueness (injectivity of $\Bound$). The boundary map $\Bound$ may or may not be a global diffeomorphism.}
\end{description}
\medskip

The local theory part of the proof is essentially linear analysis, dealing with linear elliptic boundary value problems (possibly degenerate at the boundary). The compactness issue is usually more subtle and depends crucially on the nonlinear structure of the equations. The degree calculation is also global. For a detailed study of related but more complicated (\ie,~tensor-type) boundary value problems for $AdS$ Einstein metrics (with Euclidean signature) using the method above, see~\cite{A}. 

A simpler version of this process, known as the ``method of continuity'' in elliptic PDE, is sometimes employed to prove similar global existence (and uniqueness) results. For instance, the solution of the Calabi conjecture uses the continuity method \cite{Y}. However, it is  doubtful that this method can be used to handle the equations treated here. 

\subsection{$1/2$ BPS M5 brane flows}\label{proofM5N=2}

Our first application of the process described above is to equation \eqref{heaveng},
\be \label{5.5}
\D \nonlin + 2 + (e^{\nonlin})'' = e^{\nonlin}~,
\ee
with $M=\RR\times\cC$, where $(\cC, \g)$ is a compact Riemann surface of genus ${\bf g} >1$ with fixed metric $\g$ of constant scalar curvature $R=-2$. We denote by $\D$ the Laplacian with respect to $\g$, and a prime denotes differentiation with respect to the $\RR$-coordinate $\rho$. Compared to Section \ref{sec:M5s} we have fixed the normalization $m = 1$. 

The standard solution $\nonlinss$ is given by \eqref{MNSolution} with the constant of integration set to zero,
\be
e^{\nonlinss} = 2 + e^{\r}~.
\ee
The Dirichlet problem in question is then the solvability of \eqref{5.5} for functions $\nonlin$ satisfying 
\begin{align}
\label{5.6}
\begin{split}
e^{\nonlin-\r}\to\dnonlin~,\qquad&\r\to +\infty~,\\[0pt]
e^{\nonlin}\to2~,\qquad&\r\to-\infty~,
\end{split}
\end{align}
for any $\dnonlin\in C^{\infty}(\cC)$. These boundary conditions define the asymptotic boundary map $\Bound$.

\bigskip\medskip
\noi{\bf I - Local Theory}
\medskip\\
Consider the differential operator
\be \label{5.7}
\PDO(\nonlin) = e^{-\nonlinss}[\D \nonlin + 2 + (e^{\nonlin})'' - e^{\nonlin}]~.
\ee
The moduli space $\cM$ of solutions of \eqref{5.5} satisfying boundary conditions \eqref{5.6} is given by $\PDO^{-1}(0)$. We show that the linearization $L \equiv D\PDO$ is surjective at any solution $\nonlin \in \cM$. The regular value theorem then implies that $\cM$ is a smooth manifold. 

The regular value theorem requires working in Banach (or Hilbert) spaces, although with a more technical setup one could work in Fr\'echet spaces such as $C^{\infty}$. As in Section \ref{sssec:halfBPSM5UV}, we define $\zeta = e^{-\r/2}$, and solutions $\nonlin$ of \eqref{5.5} with $C^{\infty}$ boundary value $\dnonlin$ have an asymptotic expansion at $\r \to +\infty$ of the following form (\cf~\cite{AC,Ma} for proofs of the existence of the expansion),
\be \label{5.8}
e^{-\nonlinss}e^{\nonlin} \sim \dnonlin + \nonlin_{1}\zeta + \nonlin_{2}\zeta^{2} + \nonlin_{3}\zeta^{3}  + \nonlin_{4\ell}(\log \zeta) \zeta^{4} + \nonlin_{4}\zeta^{4} + \cdots~. 
\ee
This is a polyhomogeneous expansion, in powers of $\zeta$ and $\log \zeta$, with a $\log$ term appearing at fourth order. For the moment, we work below that order, and define $\nonlin \in \w C^{3,\a}(M)$ if $e^{\nonlin-\nonlinss}$ is a $C^{3,\a}$ smooth function of $(\zeta, x, y)$, for $(x, y)$ coordinates on $\cC$ and (say) $\r > -1$. For $\r < 1$, set $\z = e^{\r/2} = e^{-|\r|/2}$ and require that $\nonlin$ is a $C^{3,\a}$ function of $(\zeta, x, y)$ with $e^{\nonlin} \to 2$ as $\r \to -\infty$. Here $C^{3,\a}$ is the H\"older function space with modulus $\a \in (0,1)$. These function spaces are used since they are well-behaved under the action of elliptic operators. We denote by $T\w C^{3,\a}(M)$ the corresponding tangent space. 

If $\lin$ is a variation of $\nonlin$, then $e^{\nonlin-\nonlinss+\e\lin} = e^{\nonlin-\nonlinss}(1 + \e{\lin} + \cdots)$, so that $\lin$ has the expansion
\be \label{5.9}
\lin \sim \dlin + \lin_{1}\z + \lin_{2}\z^{2} + \lin_{3}\z^{3} + \lin_{4\ell}(\log \z) \z^{4} + \lin_{4}\z^{4} + \cdots~.
\ee
As in Section \ref{sssec:halfBPSM5IR}, there is also an asymptotic expansion at $\rho \to -\infty$ with decay rates determined by the eigenvalues of the Laplacian $\D$. The leading order decay falls off as $e^{\r}$, so in particular $\dlin = 0$ at $\r \to-\infty$. 

The linearization $L \equiv D\PDO_{\nonlin}$ at a solution $\nonlin$ is 
\begin{align}
\label{5.10}
\begin{split}
&L:\;T \w C^{3,\a}(M)~\to~T\w C^{1,\a}(M)~,\\[5pt]
&L(\lin) = e^{-\nonlinss}[\D \lin + e^{\nonlin}\lin'' + 2(e^{\nonlin})'\lin' - (\D \nonlin + 2)\lin]~.
\end{split}
\end{align}
Then $\cM \subset \w C^{3,\a}(M)$ is a manifold if $L$ is surjective at any $\nonlin \in \cM$, \ie, the equation
\be \label{5.11}
L(\lin) = \surj~,
\ee
can be solved for arbitrary $\surj \in T\w C^{1,\a}$ with $\lin \in T\w C^{3,\a}$. 

To prove this, we first show that the operator
\be \label{5.12}
L_0: T\w C_{0}^{3,\a}(M) \to T\w C^{1,\a}(M)
\ee
is a Fredholm linear map, where the subscript denotes boundary $\dlin = 0$ at $\r \to \pm \infty$.
In the UV, \eqref{5.10} has the asymptotic form 
\be \label{5.13}
L(\lin) \sim \z^{2}\D \lin + {\tfrac{1}{4}}\dnonlin (\z^{2}\ddot \lin -3 \z\dot \lin) - \z^{2}(\D \nonlin + 2)\lin~,
\ee
where a dot denotes differentiation with respect to $\z$. This is a so-called ``totally degenerate" elliptic operator (\cf~\cite{Me}). The associated ``indicial operator" is the ODE obtained by dropping $(x,y)$-dependent and lower order terms, and is given by ${\tfrac{1}{4}}\dnonlin (\z^{2}\ddot \lin -3\z\dot \lin)$. The indicial roots are then zero and four; these are the exponents $k$ such that $\lin = \z^{k}$ solves $\z^{2}\ddot \lin - 3\z\dot \lin = 0$. Standard theory for such elliptic operators (\cf~\cite{AC,Ma,Me}) gives the Fredholm property of $L$ in \eqref{5.13}. 

Similarly, at the IR end $\r \to -\infty$, setting $\z = e^{-|\r|/2}$ and imposing the boundary condition $e^{\nonlin} \to 2$, the operator $L$ has the form
\be
L(\lin) \sim 2\D \lin +(\z^{2}\ddot \lin + \z\dot \lin) - 4\lin~.
\ee
This operator is ``totally characteristic", with indicial roots $\pm 2$; equivalently, this is a Laplace-type operator on a cylinder $\RR \times \cC$. Again, standard Fredholm theory applies for this cylindrical end (\cf~\cite{MM,Ma}).
This gives the Fredholm property for $L_{0}$ in \eqref{5.12} on either end $[\r_{0}, +\infty)$ or $(-\infty, \r_{0}]$ with, say, standard fixed Dirichlet data $\nonlin|_{\r=\r_0} = 0$ on the surface $\cC_{\r_{0}}=\{\rho_0\}\times\cC$. Taking $\r_{0} \to -\infty$ implies the Fredholm property for the full operator $L_0$. 
  
A simple computation shows that the linearization $L_{0,\sss}$ at the standard solution $\nonlinss$ is self-adjoint, with respect to the weight $e^{2\nonlinss}$. Thus the Fredholm index of $L_{0,\sss}$ is zero. This holds then for all linearizations $L_0$ in \eqref{5.12}, by invariance of the Fredholm index under deformation. 

The arguments above prove that with zero boundary values for $\lin$, \eqref{5.11} can be solved for $\surj$ in a space of finite codimension. Now let the boundary values $\dlin$ in \eqref{5.9} range over all of $C^{3,\a}(\cC)$. We claim that \eqref{5.11} is then solvable for any $\surj$. To see this, consider the following integration by parts,
\be \label{5.14}
\int_{\cC_{(-\r,\r)}}e^{2\nonlinss}\<L(\lin), \varpi\> = \int_{\cC_{(-\r,\r)}}e^{2\nonlinss}\<\lin, L^{*}(\varpi) \> + \int_{\cC_{\r}}e^{2\nonlinss}B(\lin,\varpi)~,
\ee
where $\cC_{(-\r,\r)} = (-\r,\r)\times \cC$, and $L^{*}$ is the adjoint operator (with respect to the weight $e^{2\nonlinss}$) given by\footnote{It is easily checked that $L = L^{*}$ at $\nonlin = \nonlinss$.}
\be
L^{*}(\varpi) = e^{-\nonlinss}\D \varpi +e^{-2\nonlinss}(e^{\nonlin+\nonlinss}\varpi)'' - 2e^{-2\nonlinss}((e^{\nonlin})'e^{\nonlinss}\varpi)' - e^{-\nonlinss}(\D \nonlin + 2)\varpi~.
\ee
The boundary term $B(\lin,\varpi)$ is a first order differential operator on $\lin$, $\varpi$. Now if $L$ in \eqref{5.11} is not surjective, then there exists $\varpi$ which is $L^{2}$-orthogonal to ${\rm im} L$, and hence the left-hand side of \eqref{5.14} vanishes, for all choices of $\lin$. Since $\lin$ is arbitrary, this implies 
\be \label{5.15}
L^{*}(\varpi) = 0~,
\ee
and moreover, letting $\r \to +\infty$, 
\be \label{5.16}
\z^{4}\varpi \to 0 \quad  {\rm as} \quad \r \to +\infty~,
\ee
for $\z$ as in \eqref{5.8}. The operator $L^{*}$ is elliptic, and \eqref{5.16} implies that both Dirichlet and Neumann boundary data, \ie,~the full Cauchy data, for $\varpi$ vanish at $\r \to +\infty$. All terms in the formal expansion \eqref{5.9} for $\varpi$ vanish, \cf~the discussion below. In such situations, a standard unique continuation theorem for scalar elliptic PDE (\cf~\cite{C}) implies that $\varpi = 0$, and hence $L$ is in fact 
surjective.  

This proves that the moduli space $\cM \subset \w C^{3,\a}(M)$ is a smooth Banach manifold. Clearly,
\be
T\cM = {\rm ker}\left(L\right)~.
\ee
Also $D\Bound(\lin) = \dlin$, where $L(\lin) = 0$ and with $\dlin$ as in \eqref{5.9}. The fact that the boundary map $\Bound$ is also Fredholm follows by standard linearity from the Fredholm property of $L$ in \eqref{5.12}. Briefly, \eqref{5.12} implies that one can solve $L(\lin) = \surj$, for arbitrary $\surj$ in a space $\surjset$ of finite codimension, with boundary value $\dlin = 0$ at $\r \to +\infty$. Choose now an arbitrary boundary value $\dlin$ and extend $\dlin$ to a smooth function $\dlinext$ on $M$. Then $L(\dlinext) = g$, for some function $g$, and up to a finite indeterminacy, $g \in \surjset$. For such $g \in \surjset$, one may solve $L(\lin) = g$ with $\dlin = 0$. Then $\surj = \dlinext - \lin$ solves $L(\surj) = 0$, with boundary value $\dlin$. This shows that $D\Bound$ has finite-dimensional cokernel. The proof that the range of $D\Bound$ is closed follows from elliptic regularity results. It also follows from the fact that ${\rm ind}( L_{0}) = 0$ that ${\rm ind}( D\Bound) = 0$. 

Using the boundary regularity results of \cite{AC,Ma} for the existence of the expansion \eqref{5.8}, it follows from the analysis above that the space $\cM_{\infty}$ of solutions which have smooth polyhomogeneous $C^{\infty}$ expansions is a smooth Fr\'echet manifold, with $\Bound$ a Fredholm map to $C^{\infty}(\cC)$. Thus, solutions $\lin$ to $L(\lin) = 0$ with Dirichlet boundary value $\dlin$ in $C^{\infty}(\cC)$ have the expansion \eqref{5.9} 
\be
\lin \sim  \dlin + \lin_{1}\z+ \lin_{2}\z^{2} + \lin_{3}\z^{3} + \lin_{4\ell}(\log \z) \z^{4} + \lin_{4}\z^{4} + \cdots~.
\ee
The coefficients $\dlin$, $\lin_{4}$ are the ``formally undetermined coefficients" (Dirichlet and Neumann boundary data), corresponding formally to ``source" and ``vev" perturbations. All other coefficients are determined inductively from these two. The same holds at the nonlinear level \eqref{5.8}. 

Although the Dirichlet and Neumann terms $\dlin$ and $\lin_{4}$ above are formally undetermined, one is determined globally by the other via the Dirichlet-to-Neumann map (or its inverse). Thus, specifying $\dlin$ at $\r \to +\infty$ together with the prescription $\lin \to 0$ at $\r \to -\infty$ gives (generally) a unique solution to the linearized problem $L(\lin) = 0$ with this boundary data. The resulting solution $\lin$ has an asymptotic expansion \eqref{5.9} (when $\dlin$ is $C^{\infty}$) and so the $\lin_{4}$ term is (globally) determined by $\dlin$. Again, the same holds at the nonlinear level. 

\medskip\bigskip 
\noi{\bf II - Compactness}
\medskip 

The main point in proving compactness is to derive the existence of ({\it a priori}) bounds on the maximum and minimum of a solution $\nonlin$ in terms of bounds on its boundary value $\dnonlin$ at $\r \to +\infty$, \ie,~to show that $\nonlin$ is controlled by the boundary value $\dnonlin$. To obtain a lower bound, for instance, note that the evaluation of \eqref{5.5} at any interior minimum point of $\nonlin$ implies that $e^{\nonlin} > 2$. Since $e^{\nonlin} \geq 2$ at $\r \to \pm \infty$, it follows that 
\be
e^{\nonlin} > 2
\ee
holds everywhere on $M$. Hence $\nonlin$ is uniformly bounded below. 

To obtain an {\it a priori} upper bound, multiply \eqref{5.5} by an arbitrary function $b = b(\r)$. Then
\be \label{5.17}
\D(b\,\nonlin) + b(e^{\nonlin})'' = b(e^{\nonlin} - 2)~.
\ee
Now choose $b$ so that 
\be \label{5.19}
b + b'' - 2\tfrac{(b')^{2}}{b} = 0~.
\ee
For such $b$, \eqref{5.17} can be rewritten as
\be \label{5.20}
\D (b\,\nonlin) + (b(e^{\nonlin}-2))'' -2(\log b)'(b(e^{\nonlin}-2))'  = 0~.
\ee
At an interior maximum of $b(e^{\nonlin}-2)$, the last term vanishes while the middle term is negative. Since $b = b(\r)$, a maximum of $b(e^{\nonlin}-2)$ occurs only at a maximum of $\nonlin$ on $\cC_{\r}$ for some $\rho$, so the first term is negative as well. Thus, by the maximum principle, there are no interior maxima of $b(e^{\nonlin}-2)$. Exactly the same discussion holds for minima in place of maxima. 

Now there are several solutions of \eqref{5.19}. First, let
\be \label{5.21}
b =(\cosh \r)^{-1}~.
\ee   
Then $b \sim e^{-\r}$ at $\r \to +\infty$, so that $b(e^{\nonlin}-2) \to \dnonlin$ as $\r \to +\infty$. Also $b \to 0$ at $\r \to -\infty$, so $b(e^{\nonlin}-2) \to 0$ as $\r \to -\infty$. It then follows from the above that 
\be \label{5.22}
0 < b(e^{\nonlin}-2) \leq \max \dnonlin~,
\ee
on all of $M$. This is the main {\it a priori} estimate. The boundary data $\dnonlin$ controls the pointwise size ($L^{\infty}$ norm) of any solution $\nonlin$ asymptotic to $\dnonlin$. Using the asymptotic expansion (3.39) for $\r \to -\infty$ and the test function $b = e^{-\r}$ in place of \eqref{5.21}, a similar argument shows that \eqref{5.22} may be improved to 
\be \label{5.23}
\min \dnonlin < e^{-\r}(e^{\nonlin}-2) < \max \dnonlin~.
\ee

Now suppose $\{\nonlin^{(i)}\}$ is a sequence of solutions of \eqref{5.5} with a fixed boundary value $\dnonlin$ at $\r \to +\infty$. By standard regularity theory for elliptic PDE, the sequence $\{\nonlin^{(i)}\}$ is compact (has convergent subsequences) if and only if it is bounded in $L^{\infty}$ \cite{GT,Mo,Ma}. This is given by \eqref{5.22} or \eqref{5.23}. The same remarks hold if $\dnonlin$ is replaced by a compact family $\dnonlin^{(i)} \to \dnonlin$. This establishes that the boundary map $\Bound$ is proper.
  
\bigskip\medskip
\noi{\bf III - Degree Calculation}
\medskip

We now prove that the standard solution $\nonlinss$ is the unique solution with boundary value $\dnonlin = 1$, and moreover that this solution is a regular point of the boundary map $\Bound$, \ie,~the linearization $D\Bound$ is an isomorphism at $\nonlinss$. This implies that 
\be \label{5.24}
\deg \, \Bound = 1~,
\ee
and hence $\Bound$ is surjective. 
 
To see uniqueness of $\nonlinss$, first note that a simple computation shows the functions
\be
b^{(c)} = (\cosh(\r-c))^{-1}
\ee
to satisfy \eqref{5.19}, for any constant $c$.  Thus, the discussion after \eqref{5.20} implies that the same maximum principle holds for $b^{(c)}(e^{\nonlin}-2)$. Since $b^{(c)}(e^{\nonlin}-2) \to 0$ at $\r \to-\infty$, it follows that the maximum of $b^{(c)}(e^{\nonlin}-2)$ occurs at $+\infty$, so
\be
0 \leq \frac{e^{\nonlin} - 2}{\cosh (\r-c)} \leq 2e^{c}~,
\ee
on $M$, for all $c$. 
Taking $c \to -\infty$ then implies that globally
\be \label{5.25}
e^{\nonlin} \leq 2+e^{\r}=e^{\nonlin_\sss}~.
\ee

On the other hand, integrating \eqref{5.5} over the level sets $\cC_{\r}$ of $\r$, and defining
\be
\nu \equiv \int_{\cC_{\r}}e^{\nonlin}~,
\ee
one finds that $2 + \nu'' = \nu$, which is solved by $\nu = 2 + c_{1}e^{\r} + c_{2}e^{-\r}$. As in \eqref{solnPsin}, the asymptotics at $\r \to -\infty$ imply that $c_{2} = 0$, while the asymptotics at $\r \to +\infty$ fix $c_{1} = \dnonlin = 1$. Combining this with \eqref{5.25}, it follows that 
\be \label{5.26}
\nonlin = \nonlinss~,
\ee
that is, the standard solution $\nonlinss$ is the unique solution with $\dnonlin = 1$. 

To show that the standard solution is a regular point of the boundary map, consider the linearization $L_{\sss}$ of $\PDO$ at $\nonlin = \nonlin_\sss$, given by
\be
L_{\sss}(\lin) = D\PDO_{\nonlin_\sss}(\lin) = e^{-\nonlinss}[\D\lin + e^{\nonlinss}\lin'' + 2(e^{\nonlinss})'\lin' - 2\lin]~.
\ee
Then $\lin$ is in the kernel of $D\Bound$ at this point if and only if 
\be \label{5.27a}
L_{\sss}(\lin)  = \D \lin + e^{\nonlinss}\lin'' + 2(e^{\nonlinss})'\lin' - 2\lin = 0~,
\ee
with $\lin \to 0$ at $\r \to \pm\infty$. It follows immediately from the maximum principle applied to \eqref{5.27a} that $\lin = 0$ on $M$. Thus $\ker D\Bound = 0$. Since the index of $D\Bound$ equals zero, $D\Bound$ is an isomorphism. In particular $\nonlinss$ is a regular point of $\Bound$ and thus \eqref{5.24} follows. 
 
Finally, note that it is not being claimed that $\ker D\Bound = 0$ at all solutions $\nonlin$; it remains unknown if $D\Bound$ is everywhere an isomorphism, \ie,~whether $\Bound$ is a diffeomorphism. This is due to the fact that the factor $(\D \nonlin + 2)$ of $\lin$ in \eqref{5.10} does not have a definite sign in general; its sign may change when the variation of $\nonlin$ over $\cC_{\r}$ is large. This prohibits the use of a maximum principle typically used to prove uniqueness of solutions. 

\subsection{$1/4$ BPS M5 brane flows}\label{proofM5N=1}

Consider now equation \eqref{M51/4Heaven} with the normalization $m = 1$:
\be \label{5.27}
\D \nonlin + 2 + (e^{\nonlin})'' = \tfrac{3}{2}e^{\nonlin} + e^{\nonlin}(\tfrac{1}{2}(\nonlin')^{2} - \nonlin')~.
\ee
Here the standard solution $\nonlinss$ is the one which depends only on $\r$, with asymptotics
\be
\begin{split}
e^{\nonlinss-\r} \to 1~,\qquad \r \to +\infty~,\\[0pt]
\nonlinss \to \log\,\tfrac{4}{3}~,\qquad \r \to -\infty~,
\end{split}
\ee
see Figure \ref{M5numericalfigure}.\footnote{A proof of the existence of $\nonlinss$ can be given using the techniques below, but we forgo this here.} The same strategy which was employed above can be applied to the Dirichlet problem for equation \eqref{5.27}.

 \bigskip\medskip
\noi{\bf I - Local Theory}  
\medskip

The analysis of the local theory is exactly the same as before and so we will be very brief. The analogous nonlinear operator $\PDO$ in this setting has the linearization
\be \label{5.28}
L(\lin) = e^{-\nonlinss}[\D \lin + e^{\nonlin}\lin'' + 2(e^{\nonlin})'\lin' - e^{\nonlin}(\nonlin' - 1)\lin' - (2 + \D \nonlin)\lin]~.
\ee
This has exactly the same structure as the linearized operator of Section \ref{proofM5N=2}; the indicial operator at the UV end is the same, with indicial roots zero and four, and the analysis carries over to give the same manifold structure and Fredholm results. 

\bigskip\medskip
\noi{\bf II - Compactness}
\medskip

The standard minimum principle for equation \eqref{5.27} implies that $\nonlin$ has no interior minima, and
\be
e^{\nonlin} \geq \tfrac{4}{3}~.
\ee
The main issue is then to obtain an upper bound on $\nonlin$ in terms of the Dirichlet boundary value $\dnonlin$ at $\r \to +\infty$. 

Multiplying \eqref{5.27} by $b = b(\r)$ and carrying out the same manipulations as before, with $b$ a solution to \eqref{5.19}, leads to
\be
\D (b\,\nonlin) + (b(e^{\nonlin}-2))'' - 2(\log b)'(b(e^{\nonlin}-2))' = \tfrac{b}{2}e^{\nonlin}(\nonlin' - 1)^{2} \geq 0~.
\ee
At a maximum of $b(e^{\nonlin}-2)$, the left-hand side is non-positive, while the right-hand side is positive. Choosing $b = (\cosh \r)^{-1}$, it follows as before that
\be
e^{\nonlin} -2 \leq 2(\max \dnonlin) \cosh \r~.
\ee
This gives the main {\it a priori} upper bound on $\nonlin$ in terms of $\dnonlin$. Via the same elliptic boundary regularity results, this suffices to establish the properness of the boundary map. 

\bigskip\medskip
\noi{\bf III - Degree Calculation}
\medskip

From \eqref{5.28}, the linearization $L$ at the standard solution $\nonlinss$ is given by 
\be
L_{\sss}(\lin) = e^{-\nonlinss}[\D\lin + e^{\nonlin}\lin'' + 2(e^{\nonlin})'\lin' - e^{\nonlin}(\nonlin' - 1)\lin' - 2\lin]~.
\ee
Then $\lin \in \ker D\Bound$ if and only if $L_{\sss}(\lin) = 0$ and $\lin \to 0$ at $\r \to \pm \infty$. Just as before, the maximum principle implies that the only solution of $L_{\sss}(\lin) = 0$ which vanishes at $\pm \infty$ is $\lin = 0$. Thus $D\Bound$ is an isomorphism, so $\nonlinss$ is a regular point of $\Bound$. 

We claim that $\nonlinss$ is the only solution of \eqref{5.27} asymptotic to $\dnonlin = 1$ at $\rho \to +\infty$ and to $\log (4/3)$ at $\r\to-\infty$. To prove this claim, let $\nonlin$ be any solution of \eqref{5.27} with these asymptotics. Then $\nonlin-\nonlinss \to 0$ at both asymptotic boundaries. Evaluating \eqref{5.27} on $\nonlin$ and $\nonlinss$ and subtracting gives
\begin{align}
\label{5.30}
\begin{split}
\D (\nonlin-\nonlinss) + (e^{\nonlinss})''w + 2(e^{\nonlinss})'w' + e^{\nonlinss}w'' &= \\[3pt]
{\tfrac{3}{2}}e^{\nonlinss}w + e^{\nonlinss}({\tfrac{1}{2}}(\nonlin')^{2} - \nonlin')w + {\tfrac{1}{2}}e^{\nonlinss}&(\nonlin' - \nonlinss')(\nonlin'+\nonlinss'-2)~,
\end{split}
\end{align}
where $w = e^{\nonlin-\nonlinss} - 1$, and $w \to 0$ as $\r \to \pm \infty$. Consider the evaluation of \eqref{5.30} at an interior maximum of $w$. On the first line, the first and fourth terms are non-positive and the third vanishes. On the second line, the third term vanishes. This implies the inequality
\be\label{5.31}
(e^{\nonlinss})''w\geq {\tfrac{3}{2}}e^{\nonlinss}w + e^{\nonlinss}({\tfrac{1}{2}}(\nonlinss')^{2} - \nonlinss')w~,
\ee
where we have utilized the equality of $\nonlin'$ and $\nonlinss'$ at a maximum point.  Using equation \eqref{5.27} for the standard solution, this can be rewritten as
\be\label{5.32}
0 \geq  2w~.
\ee
At an internal maximum, $w$ must take a value greater than its zero boundary value, so this is a contradiction. Thus there is no interior maximum of $\nonlin - \nonlinss$, so $\nonlin \leq \nonlinss$. The same argument, evaluating at an interior minimum, gives $\nonlin \geq \nonlinss$. Thus, $\nonlin = \nonlinss$, proving uniqueness. Hence again $\deg \, \Bound = 1$ and the boundary map $\Bound$ is surjective. 

\subsection{$1/4$ BPS D3 brane flows}\label{proofD3N=22}

We now address equation \eqref{D3N=22Heaven},
\be \label{5.33}
9\D\nonlin + (e^{\nonlin})'' = 18(e^{\nonlin}-1) + \tfrac{1}{2}e^{\nonlin}(\nonlin')^{2}~.
\ee
The standard solution $\nonlinss$ is the solution depending only on $\r$, with asymptotics
\begin{align}
\begin{split}
\nonlinss \to 0~,\qquad &\r \to -\infty~,\\[0pt]
e^{\nonlinss-6\r} \to 1~,\qquad &\r \to +\infty~,
\end{split}
\end{align}
see Figure \ref{D3numericalfigure}. (Again a proof of the existence of $\nonlinss$ can be given using the techniques below). We provide a brief discussion of the process described at the outset of this section as it applies to equation \eqref{5.33}.

\bigskip\medskip
\noi{\bf I - Local Theory}
\medskip

The local theory/manifold result is essentially the same as before. Calculating as in \eqref{5.28}, the linearization of $\PDO$ in this setting is   
\be \label{5.33a}
L(\lin) = e^{-\nonlinss}\left(9\D \lin + e^{\nonlin}\lin'' + (e^{\nonlin})'\lin' - (18\lin + 9\D \nonlin)\right)~.
\ee
Setting $\zeta = e^{-3\r}$, solutions of $\PDO(\nonlin) = 0$ and of $L(\lin) = 0$ have polyhomogenous expansions in powers of $\zeta$ and $\log \zeta$ at $\r \to +\infty$.

The indicial operator is $9\zeta^{2}\ddot \lin - 9\zeta\dot \lin$ with indicial roots  zero and two. Thus, the expansion of $e^{\nonlin-\nonlinss}$ is polyhomogenous in $\zeta$, with Dirichlet and Neumann data (source and vev) appearing at $\zeta$-exponent zero and two, respectively. Again, everything in Section \ref{proofM5N=2} carries over  to give the same manifold structure and Fredholm results. 

\bigskip\medskip
\noi{\bf II - Compactness}
\medskip

The same minimum principle as in Section \ref{proofM5N=1} gives
\be
e^{\nonlin} \geq 1~.
\ee
To obtain an upper bound depending only on the boundary value $\dnonlin$, the same argument as following \eqref{5.17} can be applied. Multiplying \eqref{5.33} by $b = b(\r)$ and setting $w = e^{\nonlin} - 1$ gives 
\be \label{5.34}
9\D (b\,\nonlin) + (bw)'' - 2(\log b)'(bw)'  = w(b'' + 18b - 2\tfrac{(b')^{2}}{b}) + \tfrac{1}{2}be^{\nonlin}(\nonlin')^{2}~.
\ee
At a critical point of $\log bw $ one finds
\be \label{5.35}
9\D (b\,\nonlin) + (bw)'' - 2(\log b)'(bw)'  = w(b'' + 18b - \tfrac{3}{2}\tfrac{(b')^{2}}{b} - 
\tfrac{1}{2}\tfrac{(b')^{2}}{b}e^{-\nonlin})~,
\ee
where at an interior maximum of $bw$, the left-hand side of \eqref{5.35} is negative. 

We now choose $b(\rho)$ to solve
\be \label{5.37}
b'' + 18b - \tfrac{3}{2}\tfrac{(b')^{2}}{b} - \tfrac{1}{2}\tfrac{(b')^{2}}{b}e^{-\nonlin_{int}} = 0~,
\ee
where $e^{\nonlin_{int}}(\r)$ denotes the average value of $e^\nonlin$ on $\cC_\r$.
Asymptotically, this equation assumes the form
\be \label{5.36}
b'' + 18b - \tfrac{3}{2}\tfrac{(b')^{2}}{b} = 0~,
\ee
which admits as a solution
\be
b = (\cosh 3\r)^{-2}~.
\ee
A solution $b$ of \eqref{5.37} exists which has the same asymptotics as $\cosh^{-2}(3\r)$.

At a maximum of $\nonlin$ on $\cC_{\r}$, the value of $\nonlin$ is greater than the average value on the Riemann surface, which by integrating \eqref{5.33} over $\cC$ can be shown to be equal to the value of $\nonlinss$ at $\rho$,\footnote{Strictly, this may require a shift of the radial coordinate as it appears in the solution $\nonlinss$. This does not affect the proof.}
\be
e^{\nonlin} \geq\frac{1}{{\rm Area}(\cC)} \int_{\cC_{\r}}e^{\nonlin}=e^{\nonlin_{int}}~.
\ee
Hence, at such a maximum of $\nonlin$ on $\cC_{\r}$,
\be \label{5.38}
9\D (b\,\nonlin) + (bw)'' - 2(\log b)'(bw)' > 0~.
\ee
It follows that $b\,\nonlin$ has no interior maxima, and hence 
\be
0 < b(e^{\nonlin} - 1) \leq \max \dnonlin~.
\ee
This is the main {\it a priori} upper bound on $e^{\nonlin}$. Again, by elliptic boundary regularity, this suffices to prove properness of the boundary map $\Bound$.

\bigskip\medskip
\noi{\bf III - Degree Calculation}
\medskip

The linearization $L_{\sss}$ of $\PDO$ at the standard solution is given by
\be
L_{\sss}(\lin) = e^{-\nonlinss}\left(9\D \lin + e^{\nonlinss}\lin''+(e^\nonlinss)'\lin'-18\lin\right)~.
\ee
Again, the standard maximum principle argument shows that the only solution $\lin$ to $L_{\sss}(\lin) = 0$ with $\lin \to 0$ at $\rho \to \pm \infty$ is $\lin = 0$. Thus the $L_{\sss}$ is an isomorphism, so $\nonlinss$ is a regular point of $\Bound$. 

The proof of uniqueness is also the same as in the previous cases. Let $\nonlin$ be any solution of \eqref{5.33} with the same asymptotics as the standard solution. Subtracting the two equations for $\nonlin$ and $\nonlinss$, as in \eqref{5.30}, yields
\begin{align}
 \label{5.39}
 \begin{split}
9\D (\nonlin-\nonlinss) + (e^{\nonlinss})''w + 2(e^{\nonlinss})'w' + e^{\nonlinss}w'' &= \\[3pt]
18e^{\nonlinss}w+ {\tfrac{1}{2}}e^{\nonlinss}(\nonlin')^{2}w + {\tfrac{1}{2}}&e^{\nonlinss}(\nonlin'+\nonlinss')(\nonlin' - \nonlinss')~.
\end{split}
\end{align}
Carrying out exactly the same arguments as appear following \eqref{5.30} leads to the bound 
$0 \geq  18w~$
at any interior maximum point. Since $w = e^{\nonlin-\nonlinss} - 1 > 0$ at such a point, this is a contradiction. Hence, 
\be
\nonlin\leq \nonlinss
\ee
holds everywhere. The same argument applied to any interior minimum point gives $\nonlin \geq \nonlinss$ everywhere. Hence, $\nonlin = \nonlinss$, proving uniqueness. So again, $\deg \, \Bound = 1$ and the boundary map is surjective.

\subsection{Area monotonicity}\label{sec:areas}

As we saw in the degree computation of Section \ref{proofM5N=2}, the geometric flow equations simplify nicely upon integration over $\cC$. In this section we take advantage of this simplification to prove that the area of the Riemann surface with metric
\be\label{Phimetric}
ds^2_{\cC}=y^{-2}e^\nonlin(dx^2+dy^2)~,
\ee
decreases monotonically along the fixed point flows of Section \ref{sec:theproof}. We solve the cases of $1/2$ BPS flows explicitly, while the $1/4$ BPS flows require a slightly more formal treatment.

Integrating the $\cN=2$ M5 brane flow \eqref{heaveng} over $\cC$ produces the ODE
\be
\cA''-\cA-4\pi\chi(\cC)=0~,
\ee
where $\cA=\int_{\cC}\exp(\nonlin)$ is the area of the Riemann surface with respect to \eqref{Phimetric} and $\chi(\cC)$ is its Euler character. The solution is given by
\be
\cA(\r)=c_1e^\r+c_2e^{-\r}+4\pi\chi(\cC)~.
\ee
The solution with the correct asymptotics to interpolate from the six-dimensional fixed point in the UV to the four-dimensional fixed point in the IR has $c_2=0$. Thus the area decreases monotonically until it reaches the fixed value at $\rho\to-\infty$.

The $\cN=(4,4)$ D3 flow \eqref{D3N=44Heaven} integrates to the following ODE:
\be
\cA''-4\cA'-36\pi\chi(\cC)=0~.
\ee
This admits the exact solution
\be
\cA(\r)=c_1+c_2e^{4\r}-9\pi\chi(\cC)\rho~.
\ee
The area is again monotonically decreasing with $\r$. As is expected, this solution does not approach a fixed point in the IR, but rather becomes singular at finite $\r$.\footnote{When the function $\cA$ is interpreted as the area of $\cC$, $\cA\to0$ is a singular limit. It should be noted that the supergravity metric function $g(\r,x,y)$ itself becomes singular along this flow, \cf, \cite{Maldacena:2000mw}.} Nevertheless, from the field theoretic point of view this is a physical flow. 

The flows preserving four supercharges do not simplify as nicely when integrated, and we can treat them simultaneously. Both flows, \eqref{M51/4Heaven}, \eqref{D3N=22Heaven},  are of the form
\be\label{genflow}
(e^\nonlin)''+k_0\D\nonlin+k_1(e^\nonlin)'-k_2(e^\nonlin)+k_3=k_4e^\nonlin(\nonlin')^2~,
\ee
with $k_1\geq0$ and $k_{2-4}>0$. Integrating over $\cC$ again eliminates the Laplacian term, but now there is a more complicated inhomogeneity in the differential equation for the area,
\be\label{genareaflow}
\cA''+k_1\cA'-k_2\cA-2\pi k_3\chi(\cC)=I[\nonlin]~,
\ee
where $I[\nonlin]$ is the integrated version of the right-hand side of equation \eqref{genflow} and is non-negative (vanishing only when $\nonlin'=0$ on $\cC$). The proof of Section \ref{sec:theproof} establishes the existence of uniformizing flows which solve \eqref{genflow} and for which $\cA$ diverges exponentially at $\rho\to+\infty$ and approaches a fixed value at $\rho\to-\infty$. Here we prove that $\cA(\r)$ decreases monotonically along these flows from the UV to the IR.

Note that for such a flow to be non-monotonic, it would have to either experience a local maximum at a finite value of $\rho$ or contain an inflection point at which $\cA'<0$. To see that neither of these scenarios can arise, define $\hat{\cA}=\cA+\frac{2\pi k_3}{k_2}\chi(\cC)$, in terms of which equation \eqref{genareaflow} simplifies to
\be
\hat{\cA}''+k_1\hat{\cA}'-k_2\hat{\cA}=I[\nonlin]~.
\ee
Note that the lower bounds on $e^\nonlin$ derived in Sections \ref{proofM5N=1} and \ref{proofD3N=22} imply that $\hat{\cA}>0$ for all $\rho$. The same requirements for monotonicity apply to the new function $\hat{\cA}$. For $\hat{\cA}'=0$, the positivity of $I[\nonlin]$ and $k_2$ imply that $\hat{\cA}''>0$, so this can only be a local minimum. Furthermore, at an inflection point of $\hat{\cA}$, one finds that $\hat{\cA}'>0$, so this does not affect monotonicity. Thus $\cA$ is a monotonic function of $\rho$ for the $1/4$ BPS uniformizing flows as well.

This monotonicity has a similar flavor to the monotonic behavior of the c-function used to prove the holographic c-theorem in \cite{Girardello:1998pd, Freedman:1999gp}, and it is tempting to identify $\cA(\rho)$ with a $(d-2)$-dimensional c-function. Indeed, such a measure of $(d-2)$-dimensional degrees of freedom would diverge in the UV where the theory is actually $d$-dimensional. It would be very interesting to derive more general monotonicity results for a function that captures the evolution of the number of degrees of freedom for flows between theories of different spacetime dimension.
 
\sect{Conclusions}\label{sec:conclusion}

We have initiated a program to use holographic BPS flows for supersymmetric wrapped branes to derive and study novel geometric flows. By extending the analysis of \cite{Maldacena:2000mw} to accommodate the presence of an arbitrary metric on the wrapped Riemann surface, we have derived a new class of elliptic equations which control the BPS flow of the conformal factor of said metric. These flow equations are particularly nice, and we have proved that they admit solutions which interpolate from any asymptotic metric in the ``UV'' to the constant negative curvature representative in the same conformal class in the ``IR''. In particular, this verifies of a crucial conjecture from the work of \cite{Gaiotto:2009we}.

In analogy with Wilsonian RG flow, it would be desirable to have holographic geometric flow equations formulated as initial value flows without the complicating factor of potentially unphysical boundary conditions. It may be that by a careful application of the tools of holographic renormalization, along with input from the field theory, one can find such a formulation for the restriction of the flows studied here to physical initial values. Alternatively, by approaching the problem using equations of motion instead of BPS equations, a more direct version of a holographic Wilsonian RG flow may be attainable \cite{Heemskerk:2010hk,Faulkner:2010jy}.

An obvious extension of our program is to the case of twisted compactification on supersymmetric cycles of dimension greater than two. In particular, the solutions of \cite{Acharya:2000mu,Gauntlett:2000ng} should be generalizable in the same way. It could be of great interest to derive a geometric flow on three-manifolds from M-theory in this way. The Ricci flow famously encounters singularities at finite time in many cases (\cf, \cite{Perelman:2006un}). One expects that a geometric flow emerging from M-theory will either avoid or provide a physical prescription for dealing with any finite-time singularities. This is currently under investigation in \cite{inprogress3}.

Finally, there are a number of natural generalizations of the present work within the two-dimensional setting. We have restricted our attention to backgrounds which preserve at least four supercharges because of certain technical simplifications which take place. In particular, this meant that we ignored the third natural class of wrapped branes  -- M2 branes -- because for M2 branes, flows with eight or four supersymmetries do not find an $AdS_2$ fixed point in the IR \cite{Gauntlett:2001qs}. There is also a $(1,1)$-supersymmetric twist of the D3 brane theory which we have neglected. Nevertheless, it may be interesting to study these less-supersymmetric compactifications and to understand whether the corresponding BPS flows display qualitatively different behavior. Furthermore, by carrying out the BPS flow analysis in ten or eleven dimensions, it should be possible to incorporate punctures.

\bigskip
\bigskip
\leftline{\bf Acknowledgements}
\smallskip
\noindent The authors would like to thank Tudor Dimofte, Mike Douglas, Abhijit Gadde, Jerome Gauntlett, Juan Maldacena, Martin Ro\v{c}ek, Eva Silverstein, A.~J.~Tolland, and especially Balt van Rees for helpful and informative discussions. C.B. thanks the Kavli Institute for Theoretical Physics (research supported by DARPA under Grant No. HR0011-09-1-0015 and by the NSF under Grant No. PHY05-51164) for generous hospitality while this work was being completed. N.B. is grateful for the warm hospitality at the Aspen Center for Physics (research supported by the NSF under Grant No. 1066293) in the final stages of this project.  M.T.A. is partially supported by NSF grant DMS-0905159. The work of C.B. and N.B. is supported in part by DOE grant DE-FG02-92ER-40697. The work of L.R. is supported in part by NSF grant PHY-0969739. Any opinions, findings, conclusions, or recommendations expressed in this material are those of the authors and do not necessarily reflect the views of the funding agencies.

\newpage
\appendixtitleon
\noappendicestocpagenum
\begin{appendices}
\sect{Derivation of Flow Equations}\label{app:derivations}
\renewcommand{\theequation}{A.\arabic{equation}}
\setcounter{equation}{0} 

In this appendix we provide a detailed account of the derivation of the flow equations for the $1/2$ BPS twist of the M5 brane theory, \eqref{BPSimpEqn}. We also provide a less thorough summary of the analogous derivation for the $1/4$ BPS M5 brane background \eqref{BPSM5N1Eqn} and for the $1/2$ and $1/4$ BPS D3 brane backgrounds \eqref{BPSimpD3N2Eqn}, \eqref{BPSimpD3N1Eqn}. Several equations from the main text are repeated here to keep the derivation relatively self-contained.

\subsection{M5 brane flows}\label{appssec:M5}

The starting point is the Ansatz for the seven-dimensional gauged supergravity background \eqref{M5metricAnsatz}, \eqref{M5gaugescalarAnsatz},
\begin{align}
\label{AAEq1}
\begin{split}
ds^2 &= e^{2f} (-dt^2+dz_1^2+dz_2^2+dz_3^2) + e^{2h}dr^2 + y^{-2}e^{2g}(dx^2+dy^2)~,\\[5pt]
 A^{(i)}&= A^{(i)}_x dx +A^{(i)}_y dy+A^{(i)}_r dr~, \qquad \lambda_i=\lambda_i(x,y,r)~, ~ i=1,2~.
\end{split}
\end{align}
As written, $(x,y)$ are coordinates on the upper half-plane, and to obtain a background with a compact $\cC$ factor we impose a quotient by a Fuchsian subgroup $\Gamma\subset PSL(2,\RR)$ which acts on the upper half-plane as
\be
z=x+iy\to \tilde{z}=\frac{az+b}{cz+d}~,\qquad\qquad ad-bc\neq0~.\label{AAEq2}
\ee
Accordingly, the functions $f$, $g$, and $h$ in \eqref{AAEq1} must be invariant under the action of $\Gamma$.\footnote{The constant negative curvature metric on the upper half-plane is given by $y^{-2}(dx^2+dy^2)$ and is invariant under all of $PSL(2,\RR)$. The conformal factor $e^g$ should then be independently invariant under $\Gamma$.} The supersymmetry variations for the relevant fermionic fields are given by \cite{Pernici:1984xx,Liu:1999ai},
\begin{align}
 \label{AAEq3}
\begin{split}
\delta\psi_{\mu} &= \left[ \nabla_{\mu} +m(A_{\mu}^{(1)}\Gamma^{12} + A_{\mu}^{(2)}\Gamma^{34}) + \ds\tfrac{m}{4} e^{-4(\lambda_1+\lambda_2)}\gamma_{\mu} + \ds\tfrac{1}{2}\gamma_{\mu}\gamma^{\nu} \partial_{\nu} (\lambda_1+\lambda_2) \right] \epsilon \\[3pt]
& + \ds\tfrac{1}{2}\gamma^{\nu} \left( e^{-2\lambda_1}F_{\mu\nu}^{(1)}\Gamma^{12} + e^{-2\lambda_2}F_{\mu\nu}^{(2)}\Gamma^{34} \right) \epsilon~, \\[5pt]
\delta\chi^{(1)} &= \left[ \tfrac{m}{4} (e^{2\lambda_1}- e^{-4 (\lambda_1+\lambda_2)}) - \tfrac{1}{4}\gamma^{\mu}  \partial_{\mu} (3\lambda_1+2\lambda_2) - \tfrac{1}{8}\gamma^{\mu\nu}e^{-2\lambda_1}F_{\mu\nu}^{(1)}\Gamma^{12} \right] \epsilon ~,\\[5pt]
\delta\chi^{(2)} &= \left[ \tfrac{m}{4} (e^{2\lambda_2}- e^{-4 (\lambda_1+\lambda_2)}) - \tfrac{1}{4}\gamma^{\mu}  \partial_{\mu} (2\lambda_1+3\lambda_2) - \tfrac{1}{8}\gamma^{\mu\nu}e^{-2\lambda_2}F_{\mu\nu}^{(2)}\Gamma^{34} \right] \epsilon ~. 
\end{split}
\end{align}
where the spin-1/2 fields $\chi^{(1)}$ and $\chi^{(2)}$ are certain linear combinations of the sixteen spin-1/2 fields of the maximal theory -- see \cite{Liu:1999ai} for more details.

We wish to find equations for the functions in \eqref{AAEq1} which guarantee the existence of spinors for which the above supersymmetry variations vanish. For a given twist of the boundary theory, we know that the generators of the preserved supersymmetries should have fixed transformation properties under the symmetries of the supergravity background. Specifically, consider the decomposition of a spinor according to
\be
\gamma_{\hat{x}\hat{y}}\epsilon=i\alpha\epsilon~,\quad\quad\Gamma^{12}\epsilon=i\beta_1\epsilon~,\quad\quad\Gamma^{34}\epsilon=i\beta_2\epsilon~,\quad\quad\gamma_{\hat r}\epsilon=\eta\epsilon~,
\label{spinorcharges}
\ee
with $\alpha$, $\beta_1$, $\beta_2$, $\eta=\pm1$.\footnote{The symplectic Majorana spinor $\epsilon$ is in the $\bf{4}$ of $SO(5)_c$, $\Gamma^i$ are $SO(5)_c$ gamma matrices and $\gamma_{\mu}$ are seven-dimensional spacetime gamma matrices. We use the standard notation $\gamma_{\mu_1\ldots\mu_p} = \gamma_{[\mu_1}\ldots\gamma_{\mu_p]}$ and suppress all spinor indices. Hats indicate flat indices.} Then the discrete parameters $\alpha$, $\beta_{1,2}$ are identified as the charges of the corresponding supersymmetry generators under $U(1)_{\cC}$, $U(1)_{12}$, and $U(1)_{34}$ as defined in Section \ref{sssec:M5twist}. This implies that for the supercharges preserved by the $1/2$ BPS twist, $\alpha=\beta_1$, while for those preserved by the $1/4$ BPS twist, $\alpha=\beta_1=\beta_2$. After fixing these relations, there are still four (resp. two) choices of signs that can be assigned in \eqref{spinorcharges}. However, each choice gives rise to the same equations for the background fields in the appropriate Ansatz.

In addition, the supersymmetries preserved by the flow should be those which restrict to Poincar\'e supersymmetries on the boundary at $r\to0_+$ (as opposed to superconformal symmetries). This fixes $\eta=1$ \cite{Lu:1996rhb}. Lastly, four-dimensional Poincar\'e invariance of the backgrounds implies that the spinors are constant in the $\RR^{1,3}$ directions,
\be
\partial_{t}\epsilon=\partial_{z_i}\epsilon=0~.\label{AAEq8}
\ee
We note that in contrast to the solutions studied in \cite{Maldacena:2000mw}, the present analysis allows for $\partial_{x}\epsilon\neq0$  and $\partial_{y}\epsilon\neq0$.

The conditions for the supersymmetry variations \eqref{AAEq3} to vanish are of two types. Vanishing of the variation of the dilatinos and the $(t,z_1,z_2,z_3)$ components of the gravitino impose explicit conditions on the background fields. Alternatively, vanishing variations of the $(r,x,y)$ components of the gravitino imply that the spinor $\epsilon$ solves a certain system of PDEs. Integrability of said system of PDEs imposes additional constraints on the background fields.

\subsubsection{$\cN=2$ M5 branes}\label{appsssec:M5N=2}

For the $1/2$ BPS twisted M5 brane background, we impose the additional simplification
\be
2\lambda_1+3\lambda_2=0~,\quad\quad A^{(2)}=0~,\label{AAEq9}
\ee
and define
\be
\lambda\equiv\lambda_2~,\quad\quad A~\equiv~ A^{(1)}~.\label{AAEq10}
\ee
To derive the BPS equations it is sufficient to take $\alpha=\beta_1=1$ in \eqref{spinorcharges}. Then the dilatino variations lead to the equations
\begin{align}
\label{AAEq11}
\begin{split}
& \partial_{r}\lambda + \ds\tfrac{2m}{5} e^{h-3\lambda} - \ds\tfrac{2m}{5}e^{h+2\lambda} + \ds\tfrac{2}{5}e^{h-2g+3\lambda} F_{xy}=0~,\\[5pt]
& (\partial_{x}+i \partial_{y}) \lambda +\ds\tfrac{2}{5} e^{-h+3\lambda}(F_{yr} -  i F_{xr})=0~, 
\end{split}
\end{align}
while vanishing of the gravitino variations (the $(t,z_1,z_2,z_3)$ components all produce the same condition) requires
\begin{align}
\label{AAEq13}
\begin{split}
& \partial_{r}\left(f-\tfrac{1}{2}\lambda\right) + \tfrac{m}{2} e^{h+2\lambda} =0~,\\[5pt]
& (\partial_{x}+i \partial_{y})\left(f-\tfrac{1}{2}\lambda\right) =0~. 
\end{split}
\end{align}
The differential equations for the spinor $\epsilon$ implied by the vanishing variations of the $(r,x,y)$ components of the gravitino are given by
\begin{align}
\label{AAEq16}
\begin{split}
&\partial_{r}\epsilon - \tfrac{1}{4}\left[\partial_{r}\lambda +m e^{h+2\lambda}+ i\,4m A_{r} \right] \epsilon\\[3pt]
&\qquad\qquad\qquad\qquad - \tfrac{1}{2}ye^{h-g}\left[(\partial_{x}+i \partial_{y})\left(h-\tfrac{1}{2}\lambda\right) -e^{-h+3\lambda}(F_{yr}-iF_{xr})\right] \gamma_6 \epsilon=0~, \\[5pt]
&\partial_{x}\epsilon + \tfrac{1}{2}\left[ i \left( \partial_{y} g - y^{-1}\right) +i\,4m A_x -\tfrac{1}{2} (\partial_{x}+i \partial_{y})\lambda +i e^{-h+3\lambda}F_{xr}  \right] \epsilon \\[3pt]
&\qquad\qquad\qquad\qquad+\tfrac{1}{2}y^{-1}e^{g-h}\left[ \partial_{r} \left(g-\tfrac{1}{2}\lambda\right) + \tfrac{m}{2}e^{h+2\lambda} - y^2 e^{h+3\lambda-2g} F_{xy} \right] \gamma_6 \epsilon=0~,\\[5pt]
&\partial_{y}\epsilon + \tfrac{i}{2}\left[ -  \partial_{x} g +4m A_y +\tfrac{1}{2} (\partial_{x}+i \partial_{y})\lambda + e^{-h+3\lambda}F_{yr}  \right] \epsilon \\[3pt]
&\qquad\qquad\qquad\qquad+\tfrac{i}{2}y^{-1}e^{g-h}\left[ \partial_{r} \left(g-\tfrac{1}{2}\lambda\right) + \tfrac{m}{2}e^{h+2\lambda} - y^2 e^{h+3\lambda-2g} F_{xy} \right] \gamma_6 \epsilon=0~. 
\end{split}
\end{align}
In order for these equations to admit solutions, they should be integrable and $PSL(2,\RR)$ covariant.\footnote{It is actually not quite necessary that the equations be covariant under $PSL(2,\RR)$. In principle, the flow could be covariant only with respect to the appropriate subgroup $\Gamma\subset PSL(2,\RR)$, or worse, the complex structure moduli of $\cC$ could vary along the flow. Fortunately, things turn out in the nicest possible way and everything is covariant.} Integrability imposes the following constraints on the background geometry and fields,
\begin{align}
\label{AAEq18}
\begin{split}
&\partial_{r}(g+2\lambda) + me^{h-3\lambda} - \tfrac{m}{2} e^{h+2\lambda} = 0~, \\[5pt]
&\partial_{r}\partial_{y}(g+2\lambda) + 2m F_{rx} = 0~, \\[5pt]
&\partial_{r}\partial_{x}(g+2\lambda) - 2m F_{ry} = 0~, \\[5pt]
& (\partial_x^2 + \partial_y^2)(g+2\lambda) + \tfrac{1}{y^2} - 2m F_{xy} = 0~.
\end{split}
\end{align}
These equations can be dramatically simplified and cast into a form which looks intrinsic to the geometry of the Riemann surface $\cC$. In particular, equations \eqref{AAEq18} fix $F_{rx}$, $F_{ry}$, and $F_{xy}$ in terms of $\lambda$, $f$, $h$, and $g$. Then \eqref{AAEq13} imply that
\be
f(r,x,y)- \tfrac{1}{2}\lambda(r,x,y) = F(r)~, \qquad\qquad h(r,x,y)+2 \lambda(r,x,y) = H(r)~,\label{AAEq22}
\ee
with $F(r)$ and $H(r)$ being real functions of the radial variable {\it only} which satisfy
\be
F^\prime(r)=-\tfrac{m}{2}\exp H(r)~.\label{AAEq23}
\ee 
This means that $F(r)$ is a monotonic function of $r$ and we can define a new radial variable $\rho$ according to
\be
\rho \equiv \tfrac{2}{m}F(r)~, \qquad\qquad \partial_\rho=-e^{-H(r)} \partial_r~.\label{AAEq24}
\ee
In terms of the new radial variable, the full solution to the BPS equations is determined by a solution to the following flow equations for the conformal factor $g$ on $\cC$ and the scalar $\lambda$,
\begin{align}
\label{AAEqFirst1}
\begin{split}
&\partial_\rho\lambda=-\tfrac{2m}{5}+\tfrac{2m}{5}e^{-5\lambda}+\tfrac{1}{5m}e^{\lambda-2g}\left(1+\Delta (g+2\lambda)\right) ~, \\[5pt]
&\partial_\rho g=\tfrac{3m}{10}+\tfrac{m}{5}e^{-5\lambda}-\tfrac{2}{5m}e^{\lambda-2g}\left(1+\Delta (g+2\lambda)\right)~, 
\end{split}
\end{align}
where we have introduced the Laplace operator on $\cC$ with respect to the metric of constant scalar curvature $R=-2$,
\be
\Delta \equiv y^2(\partial_{x}^2+\partial_{y}^2)~.\label{AAEq25}
\ee
While this is a vast improvement over \eqref{AAEqFirst1}, these flow equations are still rather complicated and can be simplified even further. After defining
\be
\nonlin (\rho,x,y) \equiv 2 g(\rho,x,y) + 4 \lambda(\rho,x,y)~,\label{AAEq26}
\ee
we find that
\be
e^{-5\lambda} = \tfrac{1}{2m} (m+ \partial_{\rho} \nonlin)~, \label{AAEqPreheaven}
\ee
where $\nonlin(\rho,x,y)$ is determined by the following second-order equation:
\medskip
\be
\boxed{\partial_{\rho}^2 e^{\nonlin} +\Delta\nonlin +2 - m^2 e^{\nonlin} = 0} \label{AAEqHeaven}
\ee
%

\subsubsection{$\cN=1$ M5 branes}\label{appsssec:M5N=1}

The $1/4$ BPS twist of the M5 brane theory leads to a different truncation of the seven-dimensional supergravity fields,
\be
A~\equiv~ A^{(1)}=A^{(2)}~, \qquad\qquad \phi\equiv-2\lambda_1= -2\lambda_2~,
\ee
and we consider a supersymmetry variation with $\alpha=\beta_1=\beta_2=1$ in \eqref{spinorcharges}. An analysis similar to the one performed for M5 branes with $\cN=2$ supersymmetry yields the equations for a supersymmetric background,
\bea
&&  \partial_{r}\phi +\tfrac{2m}{5} e^{h-\phi} -\tfrac{2m}{5} e^{h+4\phi} + \tfrac{2}{5}y^2 e^{h+\phi -2 g} F_{xy}=0~, \label{dilM5N1eq1}\\[5pt]
&& (\partial_{x}+i\partial_{y})\phi - \tfrac{2}{5}e^{-h+\phi}(F_{ry}-iF_{rx})=0~. \label{dilM5N1eq2}\\[5pt]
&& \partial_{r}\left(f- \phi\right) + \tfrac{m}{2} e^{h+4\phi} = 0~, \label{gravM5N1eq1}\\[5pt]
&& \partial_{x}\left(f- \phi\right) = \partial_{y}\left(f-\phi\right) = 0~, \label{gravM5N1eq2}\\[5pt]
&& \partial_{r}(g+4\phi) + 2 m e^{h-\phi} - \tfrac{3m}{2} e^{h+4\phi} =0 ~, \label{gravM5N1eq3}\\[5pt]
&& \partial_{r}\partial_{y}(g+4\phi) + 4m F_{rx} = 0~,\label{gravM5N1eq4}\\[5pt]
&& \partial_{r}\partial_{x}(g+4\phi) - 4m F_{ry} = 0~, \label{gravM5N1eq5}\\[5pt]
&& (\partial_x^2 + \partial_y^2)(g+4\phi) + y^{-2} - 4m F_{xy}=0~. \label{gravM5N1eq6}
\eea
Equations \eqref{dilM5N1eq1} and \eqref{dilM5N1eq2} come from the dilatino variation,  \eqref{gravM5N1eq1} and \eqref{gravM5N1eq2} from the $(t,z_1,z_2,z_3)$ components of the gravitino variation, and \eqref{gravM5N1eq3}--\eqref{gravM5N1eq6} are the integrability conditions for the PDEs which $\epsilon$ must solve. These equations can again be reformulated as a flow intrinsic to $\cC$. The result is the following system of equations in terms of a new radial variable,
\begin{align}
\label{BPSimpM5N1Eqn}
\begin{split}
&\partial_\rho\phi=- \tfrac{2m}{5} + \tfrac{2m}{5}e^{-5\phi} +\tfrac{1}{10m}e^{-3\phi-2g}\left(1+\Delta (g+4\phi)\right) ~, \\[5pt]
&\partial_\rho g=\tfrac{m}{10} + \tfrac{2m}{5}e^{-5\phi} - \tfrac{2}{5m}e^{-3\phi-2g}\left(1+\Delta (g+4\phi)\right)~. 
\end{split}
\end{align}
The new radial variable can be defined by using \eqref{gravM5N1eq1} and \eqref{gravM5N1eq2} to show that
\be
f(r,x,y)- \phi(r,x,y)= F(r)~, \qquad\qquad h(r,x,y)+4 \phi(r,x,y) = H(r)~,\label{M5n=1Radial}
\ee
in terms of which $\rho$ is defined by
\be
\rho\equiv \tfrac{2}{m}F(r)~, \qquad\qquad \partial_\rho=-e^{-H(r)} \partial_r ~.\label{M5n=1Radial2}
\ee
One can rewrite the system of two coupled PDEs \eqref{BPSimpM5N1Eqn} as a single nonlinear second-order PDE. Defining
\be
\nonlin(\rho,x,y) \equiv 2 g(\rho,x,y) + 8\phi(\rho,x,y)~,
\ee
it follows that
\be
e^{-5\phi}=\tfrac{1}{4m}\left(3m+\partial_\rho\nonlin\right)~,\label{M5n=1secondorderscalar}
\ee
where $\nonlin(\rho,x,y)$ solves the following elliptic PDE:
\medskip
\be
\boxed{\Delta\nonlin + \partial_{\rho}^2 e^{\nonlin} -e^{\nonlin}\left( \tfrac{1}{2}(\partial_{\rho}\nonlin)^2 -m \partial_{\rho}\nonlin \right) +2 - \tfrac{3m^2}{2} e^{\nonlin}=0}
\label{M5N=1Heaven}
\ee
%

\subsection{D3 brane flows}\label{appssec:D3}

The Ansatz for the twisted D3 brane solutions is analogous to the one for the twisted M5 solutions. The five-dimensional metric, the three Abelian gauge fields and two real scalars take the form
\begin{align}
\begin{split}
&ds^2 = e^{2f} (-dt^2+dz^2) + e^{2h}dr^2 + y^{-2}e^{2g} (dx^2+dy^2)~, \\[5pt]
&A^{I}= A^{I}_x dx +A^{I}_y dy+A^{I}_r dr~,\quad\quad I=1,2,3~,\\[5pt]
&\phi_1(x,y,r)~,\quad \quad \phi_2(x,y,r)~.
\end{split}
\end{align}
The coordinates $(x,y)$ are again coordinates on the upper half-plane with a quotient by a discrete subgroup of $PSL(2,\mathbb{R})$ imposed. All background fields must be invariant under the action of this discrete group. The supersymmetry transformation of the fermionic fields of the supergravity are (see \cite{Behrndt:1998ns} and Appendix A of \cite{Maldacena:2000mw} for more details),
\begin{align}
\label{AAgravitinoD3}
\begin{split}
\delta\psi_{\mu} &= \left[ \nabla_{\mu} +\tfrac{i}{8} X_{I}(\gamma_{\mu}^{\nu\rho} - 4 \delta_{\mu}^{\nu} \gamma^{\rho}) F^{I}_{\nu\rho} + \ds\tfrac{1}{2} X^{I}V_{I} \gamma_{\mu} - \tfrac{3i}{2} V_{I}A^{I}_{\mu} \right] \epsilon~, \\[5pt]
\delta\chi_{(j)} &= \left[ \tfrac{3}{8} (\partial_{\phi_j} X_{I} )F^{I}_{\mu\nu} \gamma^{\mu\nu} + \ds\tfrac{3i}{2} V_{I}\partial_{\phi_j} X^{I} - \ds\tfrac{i}{4} \delta_{jk} \partial_{\mu}\phi_k \gamma^{\mu} \right] \epsilon~, \qquad j=1,2 ~,
\end{split}
\end{align}
where we have defined
\begin{align}
\begin{split}
& X^{1} \equiv e^{-\frac{\phi_1}{\sqrt{6}}-\frac{\phi_2}{\sqrt{2}}}~, \qquad\qquad X^{2} \equiv e^{-\frac{\phi_1}{\sqrt{6}}+\frac{\phi_2}{\sqrt{2}}}~, \qquad\qquad X^{3} \equiv e^{\frac{2\phi_1}{\sqrt{6}}}~, \\[5pt]
& V_{I}=\ds\tfrac{1}{3}~, \qquad\qquad\qquad\quad\,\, X_{I} = \tfrac{1}{3} (X^{I})^{-1}~.
\end{split}
\end{align}
Since we are using an $\cN=2$ truncation of the full gauged supergravity, only a fraction of the maximal possible supersymmetry is visible. The spinors in \eqref{AAgravitinoD3}  correspond to the $(\frac{1}{2},\frac{1}{2},\frac{1}{2})$ component of the decomposition \eqref{D3spinordecomp}. In the language of this truncation the desired solutions preserve two real supercharges. In order for these to be the supersymmetries preserved by the twisted field theory, the spinors should obey the following constraints\footnote{$\gamma_{\mu}$ are the five-dimensional gamma matrices and we suppress spinor indices.}
\be
\gamma_{\hat r}\epsilon= \epsilon~, \qquad \gamma_{\hat{x}\hat{y}} \epsilon= -i  \epsilon~, \qquad \partial_{t}\epsilon=\partial_{z_i}\epsilon=0~.
\ee
Note that the radius of $AdS_5$ is fixed to one and that we allow $\partial_{x}\epsilon\neq0$  and $\partial_{y}\epsilon\neq0$.

\subsubsection{$\cN=(4,4)$ D3 branes}\label{appsssec:D3N=44}

For BPS solutions that preserve half of the maximum supersymmetry one should set
\be
\phi_2=0~, \qquad \alpha \equiv \ds\tfrac{1}{\sqrt{6}}\phi_1~, \qquad A^{(1)} = A^{(2)}=0~, \qquad A~\equiv~ A^{(3)}~.
\ee
With this simplification the analysis of the supersymmetry constraint is very similar to the case of $\cN=2$ M5 branes. First we impose the vanishing of the dilatino variations in \eqref{AAgravitinoD3}, which leads to the following differential equations
\begin{align}
\label{BPSD3N2eq7}
\begin{split}
& \partial_{r}\alpha +\ds\tfrac{2}{3} e^{h-\alpha} - \ds\tfrac{2}{3} e^{h+2\alpha} - \ds\tfrac{1}{3} y^2 e^{h- 2\alpha -2 g} F_{xy}=0~, \\[5pt]
& \partial_{x}\alpha + \ds\tfrac{1}{3} e^{-h - 2\alpha} F_{ry}=0~, \qquad\qquad \partial_{y}\alpha - \ds\tfrac{1}{3}e^{-h - 2\alpha} F_{rx}=0~.
\end{split}
\end{align}
The vanishing of the $(t,z)$ components of the gravitino variation in \eqref{AAgravitinoD3}, implies
\begin{align}
\label{BPSD3N2eq2}
\begin{split}
& \partial_{r}\left(f + \ds\tfrac{1}{2}\alpha\right) +  e^{h - \alpha} = 0~, \\[5pt]
& \partial_{x}\left(f + \ds\tfrac{1}{2}\alpha\right) = \partial_{y}\left(f + \ds\tfrac{1}{2}\alpha\right) = 0~. 
\end{split}
\end{align}
As in the case of $\cN=2$ M5 branes, the $(r,x,y)$ components of the gravitino variation lead to PDEs which should be satisfied by the spinor $\epsilon$. Integrability of this system of equations requires that the following constraints be satisfied by the background functions,
\begin{align}
\label{BPSD3N2eq1}
\begin{split}
&\partial_{r}(g - \alpha) + e^{h + 2\alpha} =0 ~, \\[5pt]
&\partial_{r}\partial_{y}(g - \alpha) + F_{rx} = 0~,\\[5pt]
&\partial_{r}\partial_{x}(g - \alpha) - F_{ry} = 0~, \\[5pt]
& (\partial_x^2 + \partial_y^2)(g - \alpha) + y^{-2} - F_{xy}=0~. 
\end{split}
\end{align}
One can simplify the system of BPS equations and reduce it to a system of two coupled PDEs intrinsic to $\cC$
\begin{align}
\label{app:BPSimpD3N2Eqn}
\begin{split}
&\partial_\rho\alpha= 2 - 2e^{3\alpha} - e^{-\alpha-2g} (1+\Delta(g - \alpha))~,\\[5pt]
&\partial_\rho g = 2 + e^{3\alpha} - e^{-\alpha-2g} (1+\Delta(g - \alpha))~.
\end{split}
\end{align}
To derive this system we have utilized a new radial variable
\be
\rho~\equiv~ \ds\tfrac{1}{3}F(r)~, \qquad\qquad \partial_\rho=-3\, e^{-H(r)} \partial_{r} ~,
\ee
where we have used
\be
f(r,x,y) + \ds\tfrac{1}{2}\alpha(r,x,y) = F(r)~, \qquad\qquad h(r,x,y) - \alpha(r,x,y) = H(r)~.
\ee
One can find a further simplification of equations \eqref{app:BPSimpD3N2Eqn} and after defining 
\be
\nonlin(\rho,x,y) \equiv 2 g(\rho,x,y) - 2 \alpha(\rho,x,y) ~,
\ee
reduce them to a single PDE that governs the flow:
\medskip
\be
\boxed{\partial_{\rho}^2e^{\nonlin} -6 \partial_{\rho}e^{\nonlin} + 9 \Delta\nonlin +18=0}
\label{AAD344Heaven}
\ee
%
\subsubsection{$\cN=(2,2)$ D3 branes}\label{appsssec:D3N=22}

To get a BPS flow that preserves a quarter of the maximal supersymmetry we set
\be
\phi_2=0~, \qquad \alpha \equiv \ds\tfrac{1}{\sqrt{6}}\phi_1~,  \qquad A~\equiv~ A^{(1)} = A^{(2)}~, \qquad A^{(3)} = 0~.
\ee
The dilatino variation yields 
\begin{align}
\label{BPSD3N1eq7}
\begin{split}
& \partial_{r}\alpha +\ds\tfrac{2}{3} e^{h-\alpha} - \ds\tfrac{2}{3} e^{h+2\alpha} + \ds\tfrac{1}{3} y^2 e^{h + \alpha -2 g} F_{xy}=0~, \\[5pt]
& \partial_{x}\alpha - \ds\tfrac{1}{3} e^{-h+\alpha} F_{ry}=0~, \qquad\qquad \partial_{y}\alpha + \ds\tfrac{1}{3}e^{-h +\alpha} F_{rx}=0~. 
\end{split}
\end{align}
The $(t,z)$ components of the gravitino variation lead to 
\begin{align}
\label{BPSD3N1eq2}
\begin{split}
& \partial_{r}\left(f - \alpha\right) +  e^{h + 2\alpha} = 0~, \\[5pt]
& \partial_{x}\left(f - \alpha\right) = \partial_{y}\left(f - \alpha\right) = 0~. 
\end{split}
\end{align}
The integrability conditions for the PDEs for the spinor $\epsilon$ coming from the $(r,x,y)$ components of the gravitino variation reduce to the following differential equations for the background fields
\begin{align}
\label{BPSD3N1eq1}
\begin{split}
&\partial_{r}(g + 2 \alpha) + 2  e^{h - \alpha} -  e^{h + 2\alpha} =0 ~, \\[5pt]
&\partial_{r}\partial_{y}(g + 2 \alpha) + 2  F_{rx} = 0~, \\[5pt]
&\partial_{r}\partial_{x}(g + 2\alpha) - 2  F_{ry} = 0~, \\[5pt]
& (\partial_x^2 + \partial_y^2)(g + 2 \alpha) +y^{-2} - 2  F_{xy}=0 ~.
\end{split}
\end{align}
Using these BPS equations one can define a new radial variable in a similar way as for the other flows above. First use
\be
f(r,x,y)- \alpha(r,x,y) = F(r)~, \qquad\qquad h(r,x,y)+2 \alpha(r,x,y) = H(r)~.
\ee
and then define the radial variable $\rho$ implicitly
\be
\rho~\equiv~\ds\tfrac{1}{3}F(r)~, \qquad\qquad \partial_\rho=-3\, e^{-H(r)} \partial_{r} ~.
\ee
With this new variable at hand one can readily derive a system of coupled PDEs intrinsic to $\cC$
\begin{align}
\begin{split}
&\partial_\rho\alpha= - 2 + 2e^{-3\alpha} +\ds\tfrac{1}{2}e^{-\alpha-2g}(1+\Delta(g+2\alpha))~,\\[5pt]
&\partial_\rho g =1 + 2e^{-3\alpha} - e^{-\alpha-2g} (1+\Delta(g+2\alpha))~.
\end{split}
\end{align}
The second-order elliptic PDE that governs the flow can be derived in terms of the new function
\be
\nonlin(\rho,x,y) \equiv 2 g(\rho,x,y) + 4 \alpha(\rho,x,y) ~.
\ee
It takes the following form:
\medskip
\be
\boxed{\partial_{\rho}^2e^{\nonlin} -\ds\tfrac{1}{2}e^{\nonlin} (\partial_{\rho}\nonlin)^2 + 9 \Delta\nonlin +18 - 18 e^{\nonlin}=0}
\label{AAD322Heaven}
\ee
%

\sect{Covariant Flow Equations}\label{app:covariant}
\renewcommand{\theequation}{B.\arabic{equation}}
\setcounter{equation}{0} 

The flow equations derived in this paper can be rewritten as covariant geometric flows. For all of the flows, the function $\nonlin$ can be interpreted as the conformal factor of an auxiliary metric on the Riemann surface $\cC$,
\be
\widetilde{ds}_{\cC}^2 = y^{-2}e^{\nonlin}(dx^2+dy^2) = e^{\Phi} (dx^2+dy^2)~.\label{ABEq1}
\ee
This metric coincides with the restriction of the gauged supergravity metric in \eqref{M5metricAnsatz} and \eqref{D3metricAnsatz} to $\cC$ in the UV, and in the IR up to a scale factor. Denoting the metric components on this Riemann surface by $g_{ij}$, the Ricci tensor is
\be
R_{ij} = - \ds\tfrac{1}{2}(\partial_{x}^2+\partial_{y}^2) \Phi ~\delta_{ij}~.\label{ABEq2}
\ee
The four second-order PDEs \eqref{AAEqHeaven}, \eqref{M5N=1Heaven}, \eqref{AAD344Heaven}, and \eqref{AAD322Heaven} can be rewritten as follows:

\begin{itemize}
\item M5 branes with $1/2$ BPS twist
\be
\partial_{\rho}^2 g_{ij} - 2 R_{ij} - m^2 g_{ij} = 0~.\label{ABEqM52}
\ee
\item M5 branes with $1/4$ BPS twist
\be
\partial_{\rho}^2 g_{ij} - 2 R_{ij} - \ds\tfrac{3m^2}{2} g_{ij} - \ds\tfrac{1}{4} \partial_{\rho}g_{i}^{k} \partial_{\rho}g_{kj} + m\partial_{\rho}g_{ij} = 0~.\label{ABEqM51}
\ee
\item D3 branes with $1/2$ BPS twist
\be
\partial_{\rho}^2 g_{ij} - 18 R_{ij} -6\partial_{\rho}g_{ij} = 0~.\label{ABEqD344}
\ee
\item D3 branes with $1/4$ BPS twist
\be
\partial_{\rho}^2 g_{ij} - 18 R_{ij} - 18 g_{ij} - \ds\tfrac{1}{4} \partial_{\rho}g_{i}^{k} \partial_{\rho}g_{kj} = 0~.\label{ABEqD322}
\ee
\end{itemize}
These covariant flow equations could form the starting point for a new, ``holographic'' proof of the uniformization theorem. Furthermore, it would be interesting to study these flow equations on higher-dimensional manifolds, and to compare the na\"ive generalization to the flows for three-manifolds which may be derived from an appropriate generalization of \cite{Acharya:2000mu,Gauntlett:2000ng} (see \cite{inprogress3}).
\end{appendices}



\begin{thebibliography}{99}
\bibitem{Perelman:2006un}
  G.~Perelman,
  ``The Entropy formula for the Ricci flow and its geometric applications,''
  [math/0211159 [math-dg]].
\end{thebibliography}
\end{document}